\documentclass[12pt,a4paper]{article}
\usepackage{graphicx}
\usepackage{times}
\usepackage{amssymb}
\usepackage{amsmath}
\usepackage{natbib}
\usepackage{mathrsfs}
\usepackage{ulem}
\usepackage{txfonts}
\usepackage{lmodern}
\usepackage{booktabs}
\usepackage{multicol}
\usepackage{cases}
\usepackage{color}

\newcommand{\aov}{\alpha_{\rm ov}}
\newcommand{\dov}{d_{\rm ov}}
\newcommand{\fov}{f_{\rm ov}}
\newcommand{\corot}{\textsc{CoRoT}}
\newcommand{\kepler}{\textit{Kepler}}

\newcommand{\plato}{\textsc{PLATO}}
\newcommand{\vaisala}{Brunt-V\"ais\"al\"a}
\newcommand{\mesa}{\textsc{MESA}}
\newcommand{\cesam}{\textsc{Cesam2k}}

\title{\bf Probing core overshooting using asteroseismology} %\footnote{A potential footnote to the title may be added here}}

\author{S\'ebastien Deheuvels$^1$\\
\vspace{0.5cm}\\
\normalsize $^1$ IRAP, Universit\'e de Toulouse, CNRS, CNES, UPS, (Toulouse), France}
\date{\mbox{}}
\begin{document}
\maketitle
\setcounter{page}{1}
\pagestyle{plain}
    \makeatletter
    \renewcommand*{\pagenumbering}[1]{%
       \gdef\thepage{\csname @#1\endcsname\c@page}%
    }
    \makeatother
\pagenumbering{arabic}

%
% WE REDEFINE THE plain LaTeX PAGESTYLE !!! 
% THIS PAGESTYLE WILL BE USED FOR THE FIRST PAGE ONLY !
% Please do not change the following lines
%
\def\bull{\vrule height .9ex width .8ex depth -.1ex}
\makeatletter
\def\ps@plain{\let\@mkboth\gobbletwo
\def\@oddhead{}\def\@oddfoot{\hfil\scriptsize\bull\quad
"How Much do we Trust Stellar Models?", held in Li\`ege (Belgium), 10-12 September 2018 \quad\bull}%
\def\@evenhead{}\let\@evenfoot\@oddfoot}
\makeatother
%
% AND DEFINE OUR MACROS FOR THE REFERENCE LIST
% I.E \beginrefer \refer and \endrefer
%
\def\beginrefer{\section*{References}%
\begin{quotation}\mbox{}\par}
\def\refer#1\par{{\setlength{\parindent}{-\leftmargin}\indent#1\par}}
\def\endrefer{\end{quotation}}
%
% BEGIN THE ABSTRACT WITH \noindent\small, ENCLOSE IT IN A GROUP
% AND BOLDFACE THE TITLE.
%
{\noindent\small{\bf Abstract:} 
Modeling properly the interface between convective cores and radiative interiors is one the most challenging and important open questions in modern stellar physics. The rapid development of asteroseismology, with the advent of space missions partly dedicated to this discipline, has provided new constraints to progress on this issue. We here give an overview of the information that can be obtained from pressure modes, gravity modes and mixed modes. We also review some of the most recent constraints obtained from space-based asteroseismology on the nature and the amount of mixing beyond convective cores.
}
\vspace{0.5cm}\\
% SPECIFY UP TO 5 KEYWORDS SEPARATED BY ' -- '

\section{Introduction}

At the occasion of the workshop "How much do we trust stellar models?" organized in Li\`ege to celebrate the 75$^{\rm th}$ birthday of Arlette Noels, I was asked to review the recent results obtained with asteroseismology to better understand the interface between convective cores and radiative interiors. This topic is both one of the most pressing open questions for stellar physics and a subject that is dear to Arlette's heart. The impact on stellar physics is clear. The mixed region associated to the convective core plays the role of a reservoir for nuclear reactions and knowing its extent is crucial to accurately model stellar evolution, in particular to estimate stellar ages. Over the last decade, the advent of spatial asteroseismology has yielded precious constraints on the size of the mixed core for stars of various masses and stages of evolution. The interpretation of these seismic data has greatly benefitted from the work of Arlette Noels and her collaborators in Li\`ege on the physical processes responsible for the extension of convective cores (overshooting, semiconvection) and their asteroseismic signature (see, e.g., \citealt{noels10}). 

%Arlette has also been at the forefront of the effort led to study the asteroseismic signature of these processes, which was a crucial step to interpret the 

Among the processes that can extend convective cores, overshooting is the most often cited. Formally, the limit of the convective core is set by the Schwarzschild criterion and it corresponds to the layer above which upward-moving convective blobs start to be braked. However, this criterion does not take into account the inertia of the ascending blobs, which can in fact \textit{overshoot} over a certain distance inside the stable region. This is expected to extend the size of the mixed core. Despite the large number of studies dedicated to this phenomenon, the details of how it operates remain very uncertain. Three physical quantities need to be determined in order to properly model core overshooting:
\begin{enumerate}
\item The distance $d_{\rm ov}$ over which chemical elements are mixed beyond the formal limit of the convective core. 
%This quantity is the most critical for stellar evolution because it determines the amount of fuel that is available for nuclear reactions. 
Theoretical studies wildly disagree on the value of $d_{\rm ov}$, with predictions ranging from 0 to several units of local pressure scale height $H_P$ (e.g., \citealt{saslaw65}, \citealt{shaviv71}, \citealt{roxburgh78}, \citealt{zahn91}). 
\item The nature of the extra-mixing beyond the convective core. Overshooting can be modeled either as an instantaneous mixing (all chemical elements being homogeneous in the overshooting region), or as a diffusive process where the turbulent velocities are generally assumed to decay exponentially in the overshoot region (\citealt{herwig00}).
\item The temperature stratification in the extra-mixing region. According to the Schwarzschild criterion, the temperature gradient should correspond to the radiative gradient ($\nabla  =\nabla_{\rm rad}$) in the overshoot region. However, the convective blobs that penetrate inside the stable regions could heat these layers and bring the temperature gradient closer to the adiabatic gradient $\nabla_{\rm ad}$. The latter case is usually referred to as \textit{penetrative convection} and by opposition, the case of an inefficient penetration that does not alter the temperature gradient ($\nabla = \nabla_{\rm rad}$ in the extra-mixed region) is referred to as \textit{non-penetrative convection}\footnote{Note that the initial terminology proposed by \cite{zahn91} was to reserve the term \textit{overshooting} for the case of an inefficient penetration. However, since the term \textit{overshooting} is widely used to refer to the general process, regardless of the temperature stratification, we prefer to use the term \textit{non-penetrative convection} instead in this review.}.
\end{enumerate}
The situation is even more complicated because other poorly-understood processes can also extend the size of convective cores, such as rotation-induced mixing (e.g. \citealt{maeder09}) or semiconvection (e.g. \citealt{langer85}). 

The combined effects of all these phenomena are generally modeled in stellar evolution codes by a simple extension of the mixed core over a distance considered as a free parameter. This distance is often referred to as the \textit{overshooting distance} and denoted as $\dov$, even though one should keep in mind that the extension of the core may in fact be caused by several distinct processes, not only core overshooting. We also use this terminology in this review. The details of how the core extension is implemented vary from one evolution code to another. The codes assuming an instantaneous mixing in the overshoot region usually take $\dov$ as a fraction $\aov$ of the pressure scale height $H_P$ at the core boundary. Core overshooting can also be implemented as a diffusive process and in this case the diffusion coefficient is generally taken as
\begin{equation}
D_{\rm ov}(r) = D_{\rm conv} \exp\left[ \frac{2(r-r_{\rm s})}{f_{\rm ov}H_P} \right]
\label{eq_ov_exp}
\end{equation}
where $r_{\rm s}$ is the radius of the Schwarzschild boundary, $D_{\rm conv}$ is the MLT diffusion coefficient some distance below $r_{\rm s}$, and $\fov$ is an adjustable parameter controlling the distance of overshooting. The temperature gradient is chosen to be either $\nabla_{\rm ad}$ (penetrative convection) or $\nabla_{\rm rad}$ (non-penetrative convection). Another important aspect is the treatment of the extension of ``small'' convective cores, for stars with masses around 1.2 $M_\odot$. The pressure scale height diverges in the center, so for small cores, the classical implementation described above generates unrealistically large core extensions that can reach the size of the core itself. Here again, evolution codes have different ways of remedying this problem. For instance, \cesam\ defines the overshooting distance as $\dov = \aov \min(H_P,r_{\rm s})$. By default, \mesa\ adopts the definition $\dov = \aov \min(H_P,r_{\rm s}/\alpha_{\rm MLT})$ where $\alpha_{\rm MLT}$ is the mixing length parameter. Considering these definitions, small convective cores can have extensions over distances that vary by a factor $\alpha_{\rm MLT}$ for the same value of $\aov$ in the two codes. Other studies chose to impose a linear dependence of $\aov$ on stellar mass in this mass range (e.g., \citealt{pietrinferni04}, \citealt{bressan12}). One should be aware of these differences, which prevent direct comparisons of overshooting efficiencies between codes that adopt different prescriptions. 

The diversity of these implementations is due to the current lack of observations that could help constrain the physical properties of the extra-mixing beyond convective cores. So far, constraints on core overshooting were obtained mainly from the modeling of eclipsing binaries (e.g. \citealt{stancliffe15}, \citealt{claret18}) and from the color-magnitude diagrams of clusters (e.g., \citealt{maeder81}, \citealt{vandenberg06}). These observational data are essentially sensitive to the distance of the extra-mixing $d_{\rm ov}$. Asteroseismology can directly probe the size of the mixed core at current age because oscillation modes are sensitive to the sharp gradient of chemical composition at this location. As will be shown in the following sections, seismic constraints can also be obtained on the chemical profile within the region of extra-mixing and on the temperature stratification, which opens the interesting prospect of testing more complex models of core overshooting.

In this review, we present a selection of the most recent seismic constraints on the physical properties of the boundaries of convective cores. Our aim is not to be exhaustive, but to give an overview of the latest developments made possible thanks to space-based asteroseismology. For this purpose, we focus on four types of stars. We start on the main sequence with results obtained for solar-like pulsators using pressure modes (Sect. \ref{sect_solarlike}) and for slowly pulsating B (SPB) stars using gravity modes (Sect \ref{sect_classical}). We then show that mixed modes can also place strong constraints on core overshooting in subgiants (Sect. \ref{sect_subgiants}) and core-helium burning giants (Sect. \ref{sect_CHeB}).

\section{Constraints from main sequence solar-like pulsators \label{sect_solarlike}}

\subsection{What constraints can we expect from pressure modes?}

\begin{figure}[h]
\centering
\includegraphics[width=10cm]{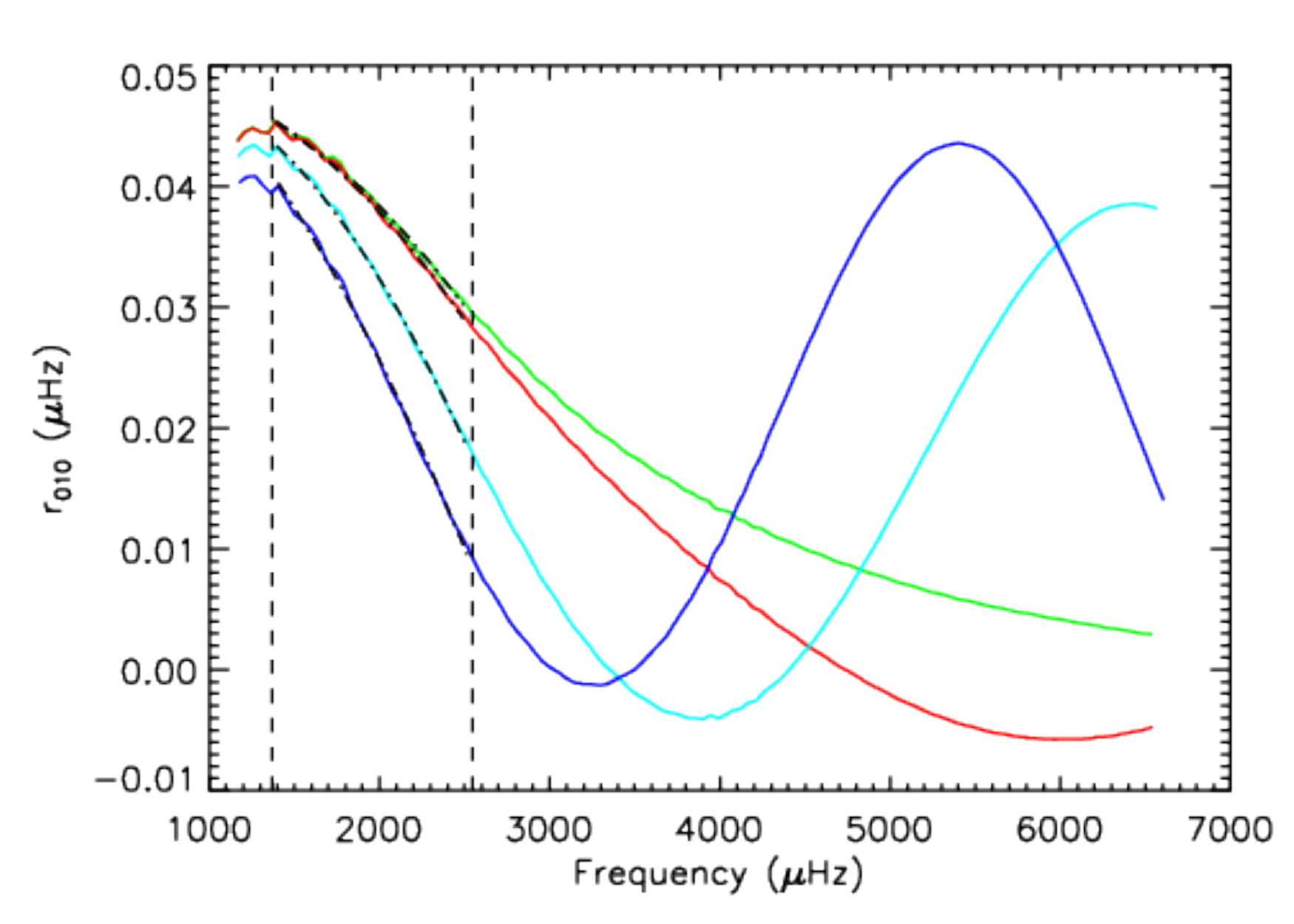}
\caption{Variations in the ratio $r_{01}$ as a function of frequency for 1.15-$M_\odot$ main sequence models with $\aov = 0$ (green), $0.1$ (red), $0.15$ (cyan), $0.2$ (blue). The vertical dashed lines indicate the frequency interval where solar-like oscillations are expected to be excited. From \cite{deheuvels15b}.
\label{fig_ratio_ov}}
\end{figure}

Pressure modes are sensitive to the region of extra-mixing beyond the convective core through its effect on the sound speed velocity $c_{\rm s}$. Assuming an ideal gas law, $c_{\rm s}^2 = \Gamma_1 \mathcal{R}T/\mu$, where $\Gamma_1$ is the adiabatic exponent, $T$ is the temperature, and $\mu$ is the mean molecular weight. At the boundary of the mixed core, a strong $\mu$-gradient develops, which creates a near discontinuity in the sound speed velocity. This generates an \textit{acoustic glitch} for pressure modes (the spatial scale of the variations in $c_{\rm s}$ is smaller than the mode wavelength), which produces a clear signature in the frequencies of these modes. It is well known that acoustic glitches generate a periodic modulation of the mode frequencies (\citealt{gough90}). The amplitude of the modulation depends on the intensity of the glitch (sharpness of the $\mu$-gradient) and the period depends on the location of the glitch (the deeper the boundary of the mixed core, the longer the period). In principle, pressure modes thus convey information about the size of the mixed core and the nature of the mixing in the overshoot region. Note that acoustic glitches are also produced by the bottom of the convective envelope (e.g., \citealt{christensen11}) and the zone of ionization of helium (e.g., \citealt{mazumdar14}, \citealt{verma14}).

Although the periodic modulation due to the acoustic glitch is present in the mode frequencies themselves, it is more convenient to use combinations of mode frequencies instead. Most studies use small differences $d_{01}$ or second differences $dd_{01}$ built with radial and dipolar modes
\begin{align}
d_{01} & = \frac{1}{2} \left( - \nu_{1,n-1} + 2\nu_{0,n} - \nu_{1,n}  \right) \\
%d_{10} & = -\frac{1}{2} \left( - \nu_{0,n} + 2\nu_{1,n} - \nu_{0,n+1}  \right),
%\end{align}
%or second differences
%\begin{align}
dd_{01} & = \frac{1}{8} \left( \nu_{0,n-1} - 4\nu_{1,n-1} + 6\nu_{0,n} - 4\nu_{1,n} + \nu_{0,n+1}  \right).
%dd_{10} & = -\frac{1}{8} \left( \nu_{1,n-1} - 4\nu_{0,n} + 6\nu_{1,n} - 4\nu_{0,n+1} + \nu_{1,n+1}  \right)
\end{align}
It has indeed been shown that these quantities are particularly sensitive to the structure of the core (e.g., \citealt{provost05}). Besides, the ratios $r_{01}$ defined as  $dd_{01}/\Delta\nu_1$, where $\Delta\nu_1$ corresponds to the large separation of dipolar modes ($\Delta\nu_{1,n} = \nu_{1,n}-\nu_{1,n-1}$), have been shown to be largely insensitive to the structure of the outer layers, which makes them almost immune to the well-known near-surface effects (\citealt{roxburgh03}).

As an illustration, Fig. \ref{fig_ratio_ov} shows the variations in the ratio $r_{01}$ with frequency for 1.15 $M_\odot$ main sequence models. Different extensions of the convective core were considered, ranging from $\aov = 0$ to $\aov = 0.2$, and models were evolved until the same age. For $\aov>0.1$, the models have a convective core at the current age and the periodic modulation caused by the edge of the core is clearly visible. It is also evident that for larger core sizes (when $\aov$ increases), the period of the oscillation decreases. Fig. \ref{fig_ratio_ov} also shows the approximate range of frequencies where p-modes are expected have detectable amplitudes. It appears that this interval is much shorter than the period of the modulation, which unfortunately prevents us from getting model-independent measurements of the size of the mixed core. However, the behavior of $r_{01}$ in the range of observed modes changes significantly as $\aov$ is varied, showing that the extent of the convective core can be determined using model-dependent analyses. In particular, the coefficients of a linear regression of $r_{01}(\nu)$ have been shown to efficiently constrain the amount of extra mixing (\citealt{deheuvels10}, \citealt{silva11}). Several analyses of this type have been recently obtained.

\subsection{Some recent results}

%\noindent \textit{HD203608:} \cite{deheuvels10} \\

\noindent \textit{HD49933:} HD49933 is an F5-type main sequence star and was the first solar-like pulsator to be observed with the \corot\ satellite. It benefitted from 180 days of nearly continuous observations and the properties of its oscillation modes were determined by \cite{benomar09b}. The identification of the degree of the detected modes initially caused problems, an ambiguity arising between the $l=1$ rotationally split modes and the overlapping $l=0$ and $l=2$ modes. This problem is now known to occur for all F-type pulsators owing to their large mode width, and several methods have been proposed to remedy this issue (e.g., \citealt{bedding10}). The mode identification for HD49933 is now robust, and \cite{goupil11} performed a modeling of the star. They found that HD49933 has a stellar mass of in the range 1.05-1.18\,$M_\odot$ and an age in the range 2.9-3.9\,Gyr. They showed that to reproduce the behavior of the observed small differences $d_{01}$, an extension of the convective core over a distance $\dov \approx 0.2 H_P$ needs to be invoked.  They also calculated models of the star including microscopic diffusion and rotationally-induced mixing using the code CESTAM (\citealt{marques13}). They found that these models fail to reproduce the slope of $d_{01}(\nu)$ and that some amount of core overshoot needs to be included to produce a good agreement with the seismic data. \\

\begin{figure}[h]
\begin{minipage}{8cm}
\centering
\includegraphics[width=8cm]{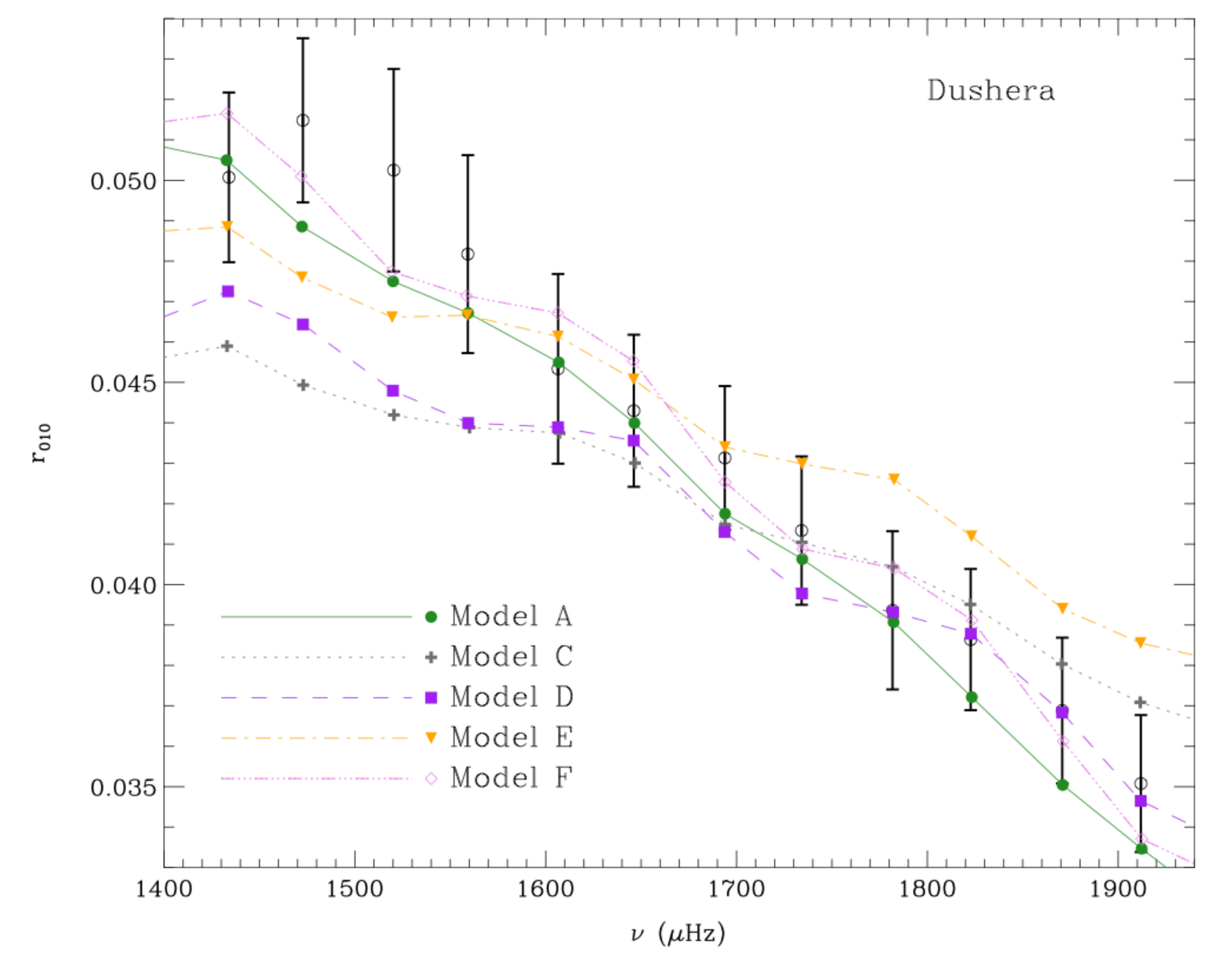}
\caption{Ratios $r_{01}$ of KIC12009504 (open circles). The colored symbols correspond to models computed with various evolutions codes and input physics (see \citealt{silva13}). \label{fig_silva13}}
\end{minipage}
\hfill
\begin{minipage}{8cm}
\centering
\includegraphics[width=8cm]{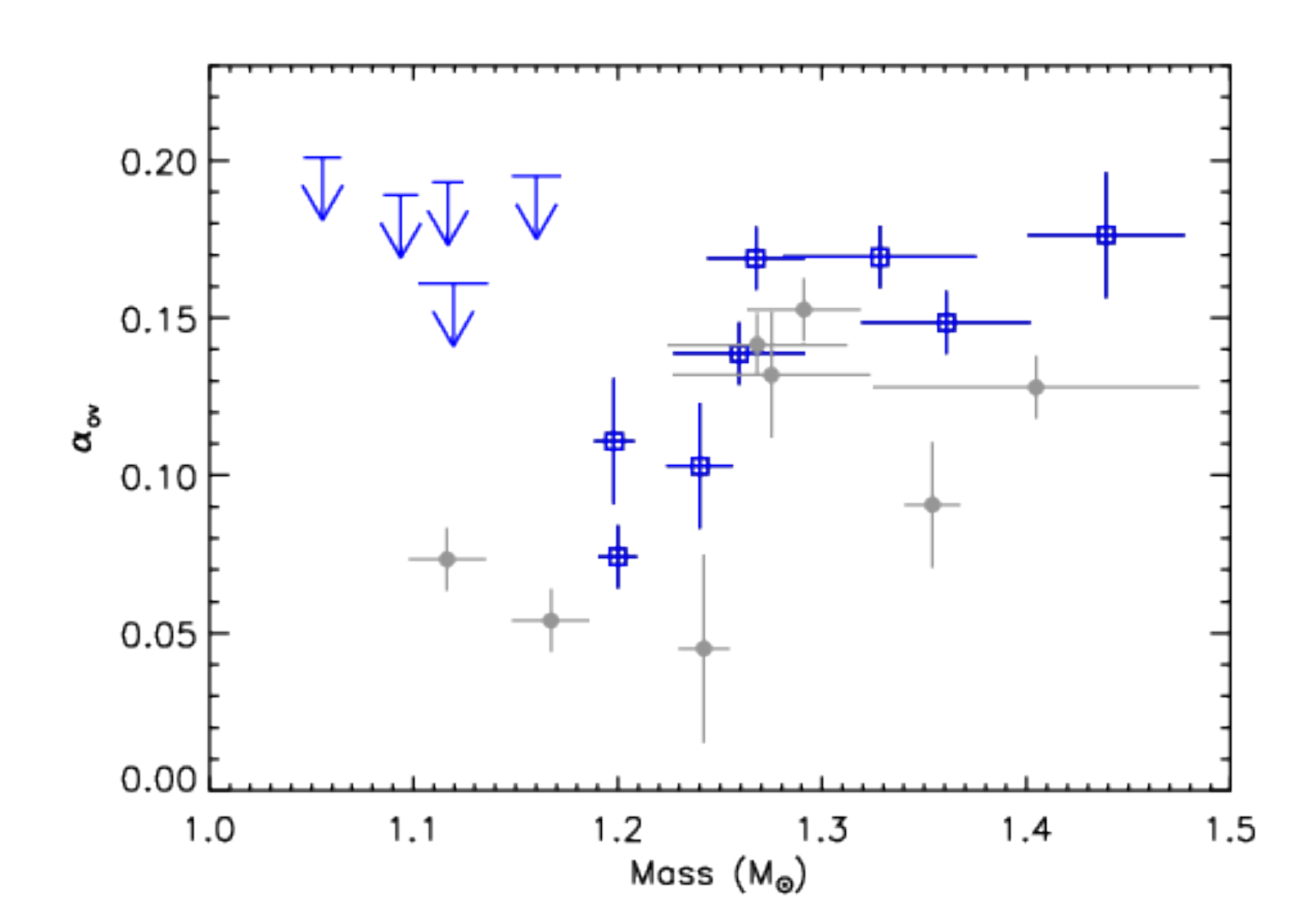}
\caption{Amount of core overshooting required for the eight stars studied by \cite{deheuvels16} plotted as a function of the stellar mass. Blue squares (resp. gray circles) indicate models computed without (resp. with) microscopic diffusion. Vertical arrows indicate upper values of $\aov$ for five other stars. \label{fig_deheuvels16}}
\end{minipage}
\end{figure}

\noindent \textit{KIC12009504 (Dushera):} The \kepler\ satellite has provided us with nearly four years of continuous observations during the nominal mission. An early analysis of the \kepler\ main sequence target KIC12009504 (dubbed Dushera) already permitted to find evidence that the star has a convective core and to place constraints on its extent (\citealt{silva13}). The authors modeled the star using nine months of \kepler\ data analyzed by \cite{appourchaux12b}. They found that the star has a stellar mass of $1.15\pm0.04\,M_\odot$, a radius of $1.39\pm0.01\,R_\odot$ and an age of $3.80\pm0.37$ Gyr. They also showed that the observed ratios $r_{01}$ could be reproduced only by models with a convective core that extends beyond the Schwarzschild boundary (see Fig. \ref{fig_silva13}). Optimal fits were obtained when the limit of the mixed core is located at an acoustic radius equal to $\sim2.4\%$ of the total acoustic radius\footnote{The acoustic radius is defined as $\tau \equiv \int_0^r \hbox{d}r/c_{\rm s}$. It corresponds to the wave travel time from the center to a radius $r$.}. \\

\noindent \textit{Dependence of the amount of overshoot with stellar mass:} As illustrated by the examples presented here, several asteroseismic studies were led on individual stars, which all reported the need for extended convective cores. It is important now to have access to consistent studies of larger samples of stars in order to better understand how the efficiency of the extra-mixing beyond convective cores depends on global stellar properties. \cite{deheuvels16} modeled 24 \kepler\ solar-like pulsators in a consistent way, using the coefficients of a $2^{\rm nd}$-order polynomial fit to the ratios $r_{01}$ to probe the mixed core. Within this sample, 10 stars were found to be already on the post-main-sequence. Among the other targets, the authors detected a convective core in eight stars and they were able to estimate the size of their mixed core, finding a good agreement with the two evolution codes \cesam\ and \mesa\ (using identical prescriptions for core overshooting). It was necessary to include significant extensions of the mixed core in all the considered targets. The optimal values of $\aov$ obtained for these eight stars are shown as a function of stellar mass in Fig. \ref{fig_deheuvels16}. As can be seen in this figure, there seems to be a tendency of core overshooting to increase with stellar mass in the considered mass range, although more data points will be required to confirm this trend. Interestingly an increase of the efficiency of core overshooting with mass was also found using constraints from double-lined eclipsing binaries by \cite{claret18}, although this result is currently debated (\citealt{constantino18}). One should also beware that the stars studied by \cite{deheuvels16} are in the range of mass where the radius of the convective core is smaller than the pressure scale height at the core edge during most of the main sequence evolution. The efficiency of the extra-mixing beyond the convective core parameterized by $\aov$ thus depends on the treatment that they adopted for ``small'' convective cores ($\dov$ redefined as $\aov r_{\rm s}$ when $H_P>r_{\rm s}$ in this study).  \\

%\subsection{Inversions}
%
%Inversions, \cite{buldgen18}

\section{Constraints from main sequence g-mode pulsators \label{sect_classical}}

%The \corot\ and \kepler\ missions have produced exquisite photometric data for g-mode classical pulsators, in particular $\gamma$ Doradus stars and slowly pulsating B (SPB) stars.

Gravity modes are expected to be excellent probes of the region of extra-mixing beyond the convective core, through their dependence on the \vaisala\ frequency $N$ (see Sect. \ref{sect_gmodes}). The \corot\ and \kepler\ missions have produced exquisite photometric data for g-mode classical pulsators, in particular slowly pulsating B (SPB) stars and $\gamma$ Doradus stars, thus providing information about core properties for stars of intermediate masses.

\subsection{What constraints can we expect from gravity modes? \label{sect_gmodes}}

High-order gravity modes (in the asymptotic regime) are expected to be equally spaced in period. The asymptotic period spacing of g modes of degree $l$ is approximately given by 
\begin{equation}
\Delta\Pi_l \approx \frac{2\pi^2}{L} \int_{r_{\rm i}}^{r_{\rm o}} \left( \frac{N}{r} \,\hbox{d}r \right)^{-1},
\label{eq_deltap}
\end{equation}
 where $L^2 = l(l+1)$ and the radii $r_{\rm i}$ and $r_{\rm o}$ are the inner and outer turning points of the g-mode cavity. The \vaisala\ frequency directly depends on the temperature stratification and the $\mu$-gradient in the g-mode cavity through the relation 
 %$N^2 = N_T^2 + N_\mu^2$, where $N_T$ and $N_\mu$ are known as the thermal part and the chemical part of the \vaisala\ frequency. They are defined as 
%\begin{equation}
%N_T^2 = \frac{g\delta}{H_P} \left( \nabla_{\rm ad} - \nabla \right) \;\; ; \;\; N_\mu^2 = \frac{g\varphi}{H_P}  \nabla_\mu 
%\end{equation}
\begin{equation}
N^2 = \frac{g\delta}{H_P} \left( \nabla_{\rm ad} - \nabla + \frac{\varphi}{\delta} \nabla_\mu \right)
\end{equation}
where $\nabla_\mu \equiv (\hbox{d}\ln\mu/\hbox{d}\ln P)$, $\delta = \left(\partial\ln\rho/\partial\ln T\right)_{P,\mu}$, and $\varphi = \left(\partial\ln\rho/\partial\ln\mu\right)_{P,T}$. 

\begin{figure}
\centering
\includegraphics[width=5.6cm]{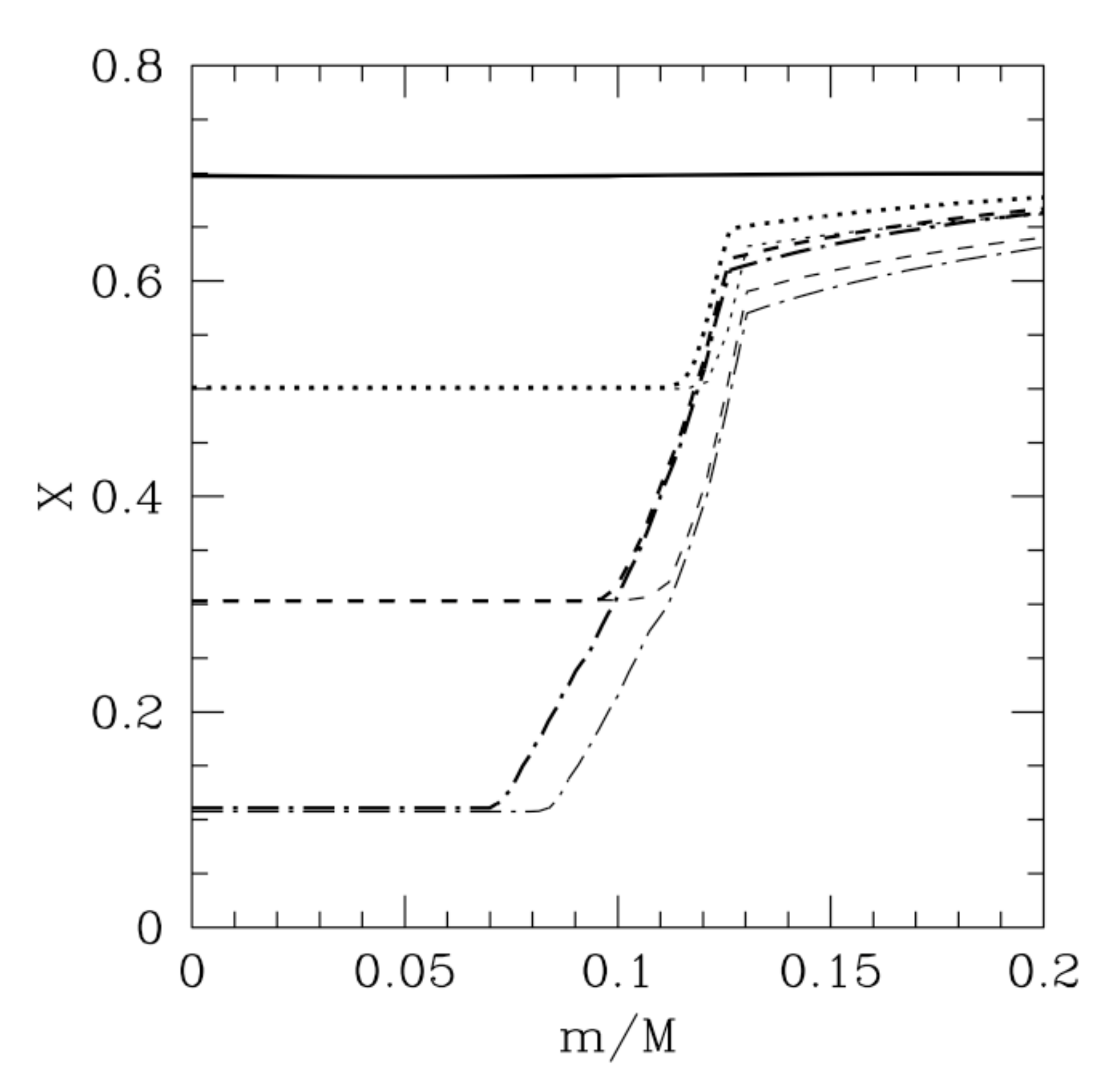}
\includegraphics[width=5.4cm]{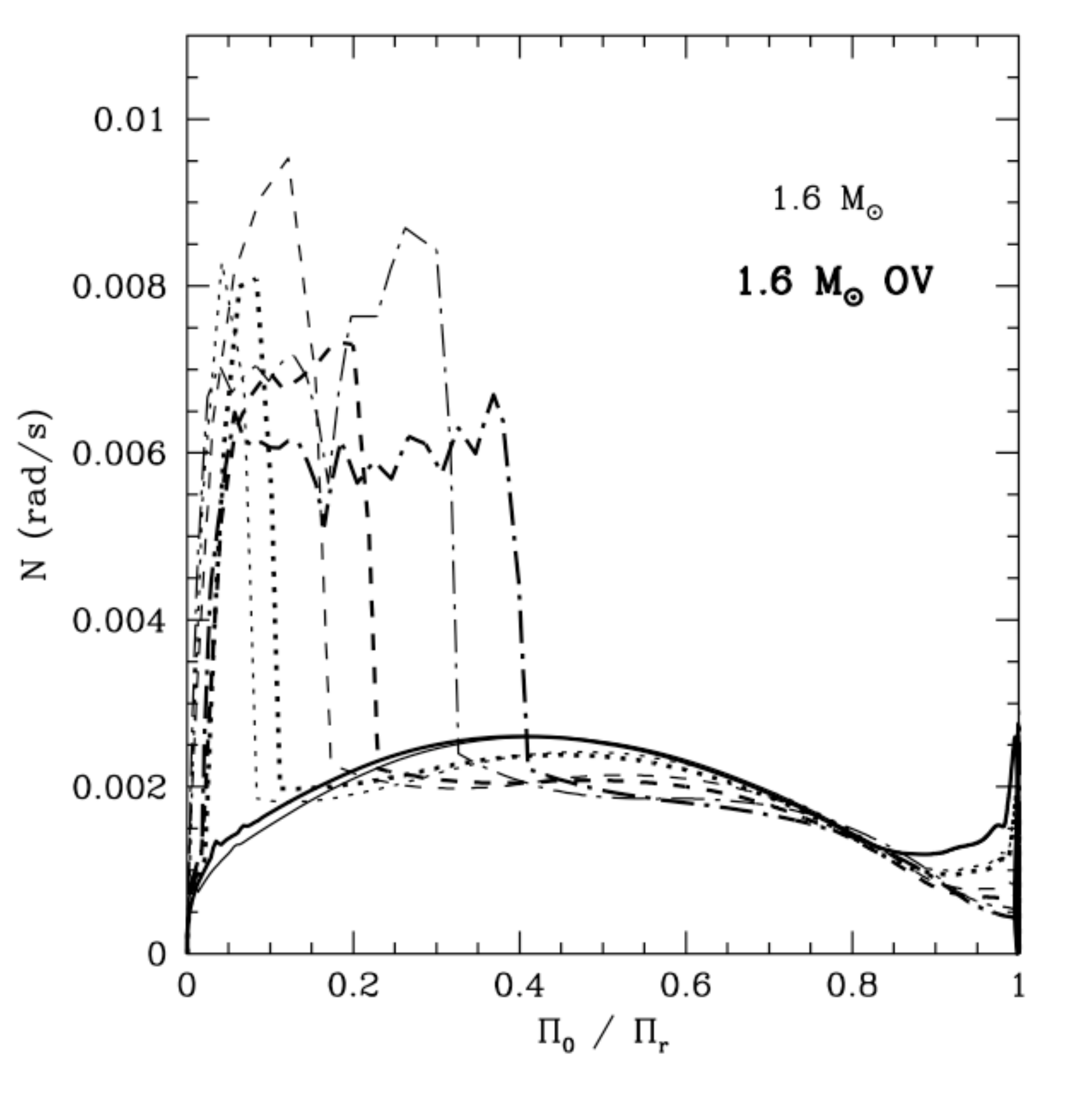}
\includegraphics[width=5.6cm]{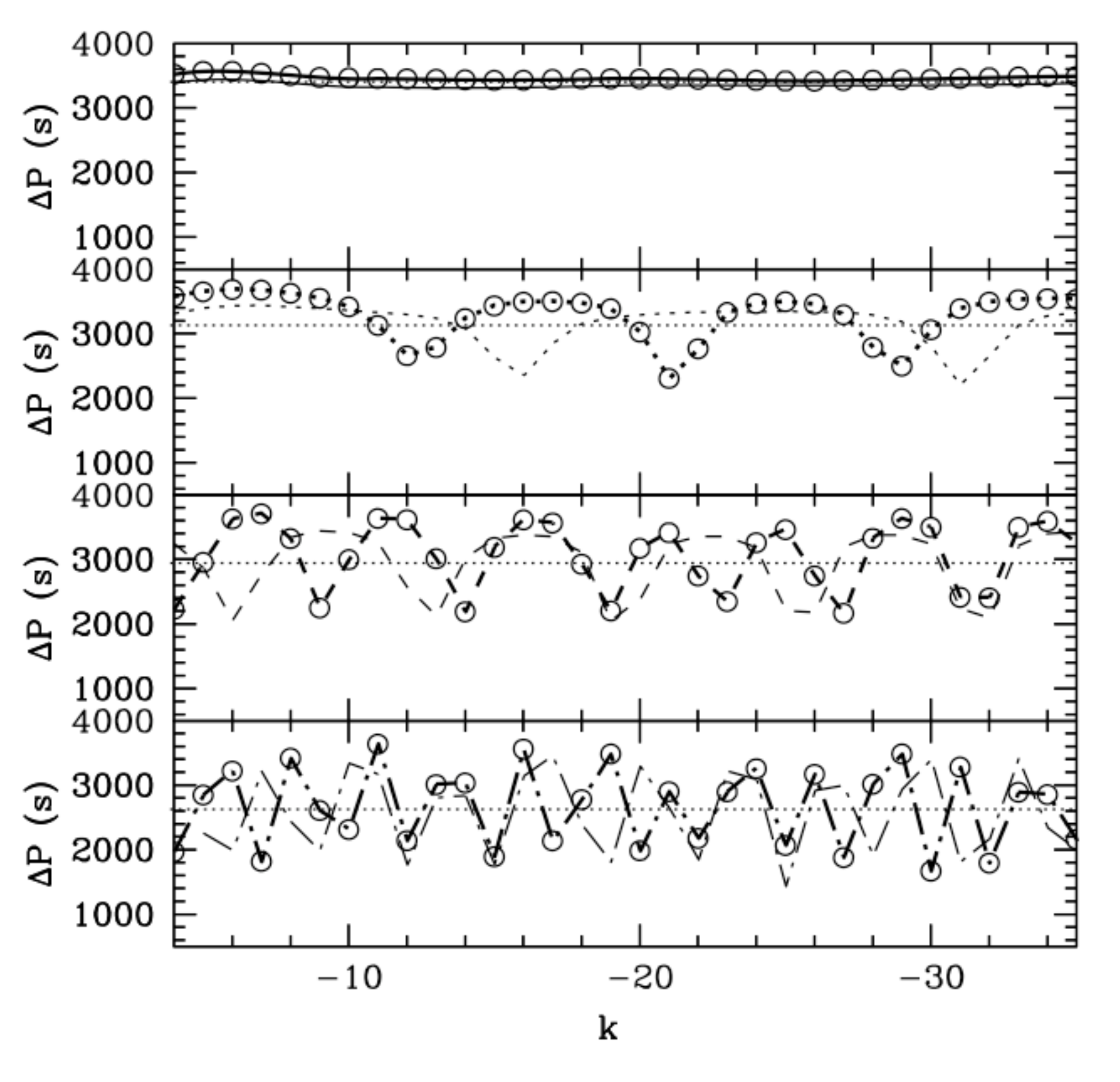}
\caption{Behavior of the hydrogen abundance profile (left), of the \vaisala\ frequency (center) and of the $l=1$ g-mode period spacing (right) in a 1.6-$M_\odot$ model computed with (thick lines) or without (thin lines) overshooting (from \citealt{miglio08}).
\label{fig_miglio08}}
\end{figure}

As stars evolve, the hydrogen content in the convective core decreases and a region of increasingly large $\mu$-gradient develops above the boundary of the core (see Fig. \ref{fig_miglio08}). This generates a \textit{buoyancy glitch} at the outer edge of the $\mu$-gradient region, where $\nabla_\mu$ varies on a length scale that is shorter than the mode wavelength. This glitch produces a periodic modulation of $\Delta\Pi_l$, whose period depends on the location $r_\mu$ of the glitch within the cavity (the deeper the glitch, the longer the period, as can be seen in the right panel of Fig. \ref{fig_miglio08}). 
%The period can indeed be expressed in terms of the radial order $n$ as
%\begin{equation}
%\Delta n = \frac{\Pi(r_\mu)}{\Pi(r_{\rm o})}
%\label{eq_period_g}
%\end{equation}
%where $\Pi(r)$ is the \textit{buoyancy radius} defined as
%\begin{equation}
%\Pi(r) = \left( \int_{r_{\rm i}}^{r} \frac{N}{r'} \,\hbox{d}r' \right)^{-1}
%\label{eq_buoyancy_rad}
%\end{equation}
%and $r_{\rm i}$ and $r_{\rm o}$ are the inner and outer turning points of the g-mode cavity. Thus, the period $\Delta n$ is given by the ratio between the buoyancy radius of the glitch and the total buoyancy radius of the cavity (the deeper the glitch, the longer the period). 
The amplitude of the modulation depends on the intensity of the glitch, i.e., on the smoothness of the chemical profile outside the convective core. For stars massive enough for the CNO cycle to dominate during their main-sequence evolution, the convective core recedes, which increases the size of the $\mu$-gradient region. As a result, the outer edge of the $\mu$-gradient region moves outwards and the period of the modulation decreases, as can be seen in the right panel of Fig. \ref{fig_miglio08}.

\begin{figure}
\begin{minipage}{6.5cm}
\centering
\includegraphics[width=6.5cm]{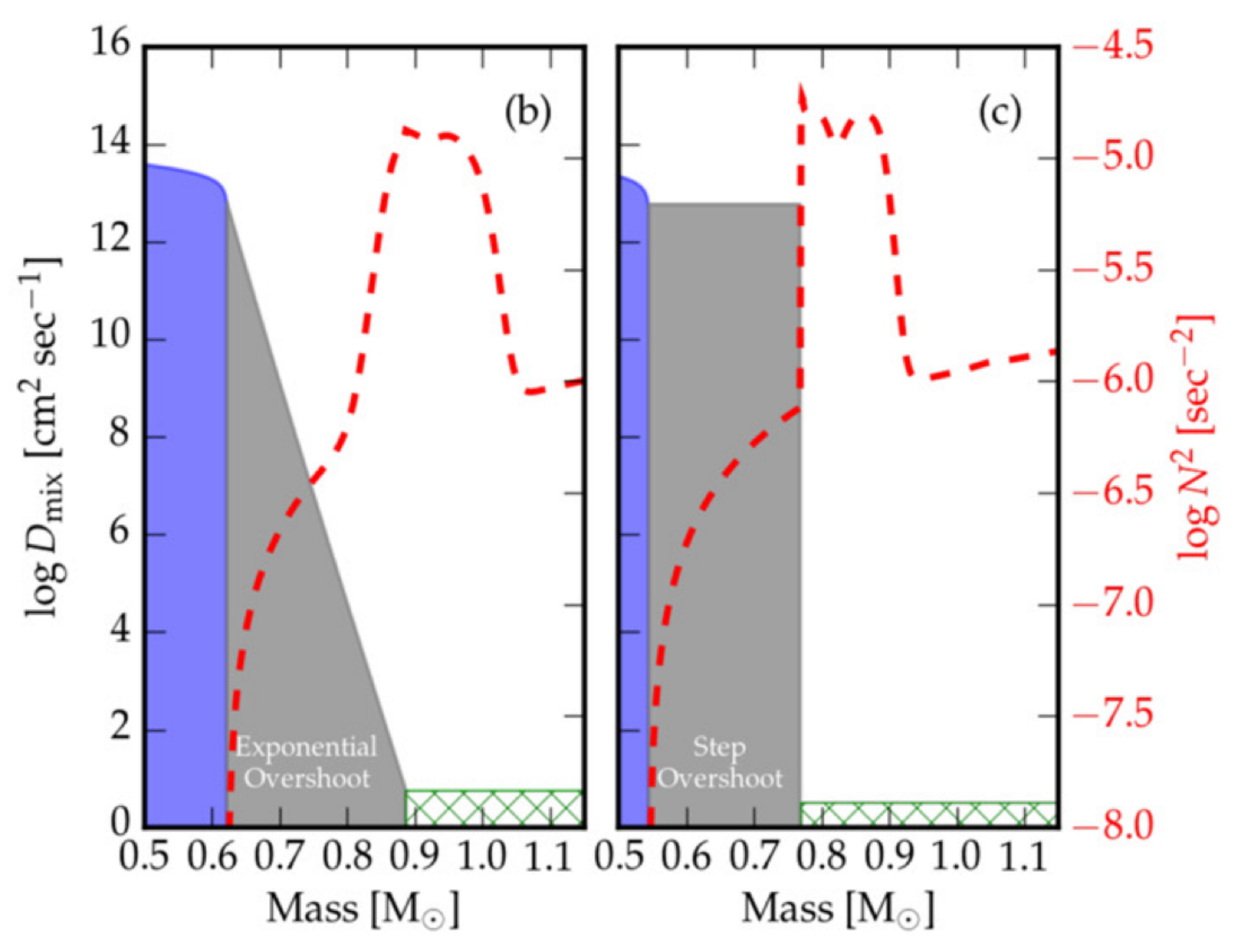}
\caption{Profile of the \vaisala\ frequency (red dashed curves) where overshooting is treated as a diffusive process (left) or as an instantaneous mixing (right) (from \citealt{moravveji16}). \label{fig_morav16}}
\end{minipage}
\hfill
\begin{minipage}{9.5cm}
\centering
\includegraphics[width=9.5cm]{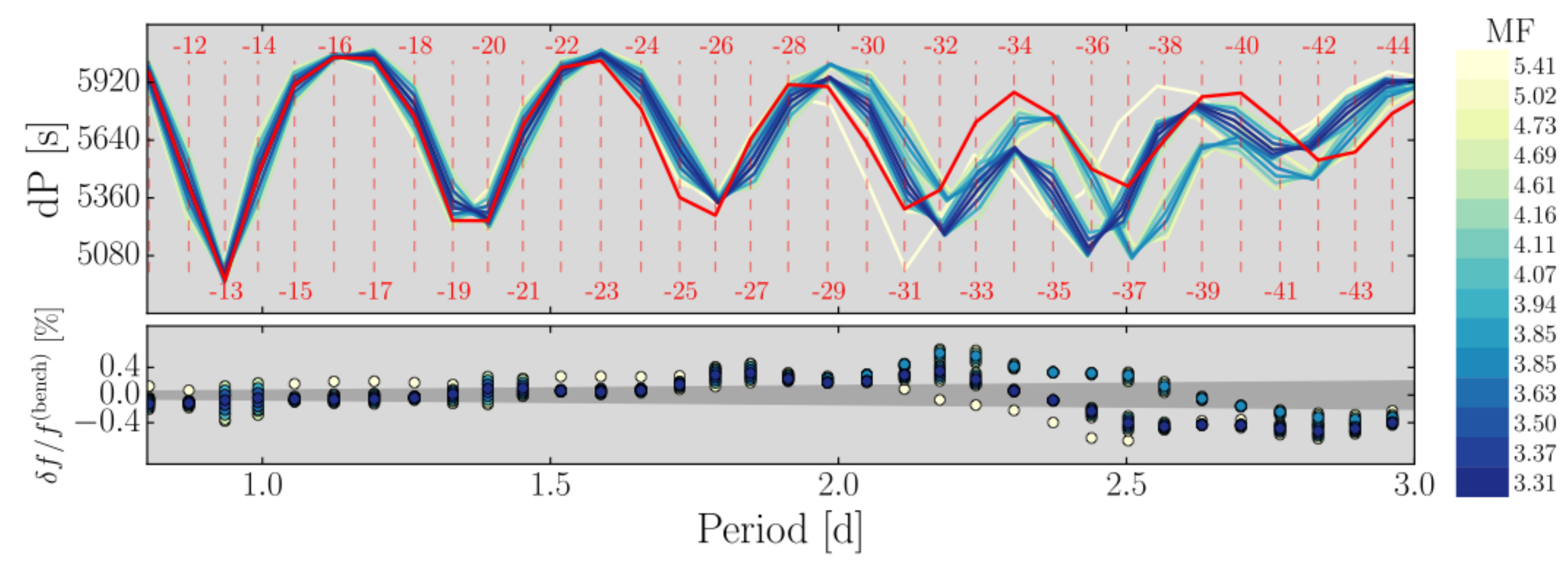}
\caption{Comparison between the period spacings of a reference model with diffusive overshoot (red curve) and the 15 best matching models of a grid computed with instantaneous mixing. The colors indicate the level of agreement with the reference model (from \citealt{pedersen18}). \label{fig_pedersen18}}
\end{minipage}
\end{figure}

The characteristics of this periodic modulation give direct constraints on the properties of the extra-mixing beyond the convective core. As shown by \cite{miglio08}, adding core overshooting to stellar models changes the size of the $\mu$-gradient region and thus modifies the period of the modulation (see Fig. \ref{fig_miglio08}). The nature of the mixing in the overshoot region can also be tested. When core overshooting is treated as a diffusive process, $\nabla_\mu$ varies more smoothly than when an instantaneous mixing is assumed and the glitch produced in the \vaisala\ frequency is less steep (see left panel of Fig. \ref{fig_morav16}). This makes a difference for g-modes with higher periods. These modes have shorter wavelengths, which eventually become smaller than the length scale of the sharp feature in $\nabla_\mu$ as mode period increases. Thus, higher-period g modes do not ``feel'' this feature as a glitch and we expect the amplitude of the periodic modulation in $\Delta\Pi_l$ to decrease as mode period increases. The situation is different for an actual discontinuity in the \vaisala\ frequency, for which all the modes have wavelengths longer than the length scale of the glitch. 

To test this quantitatively, \cite{pedersen18} calculated a reference model of 3.25 $M_\odot$ with diffusive overshooting and tried to see if its seismic content could be reproduced by models computed with an instantaneous overshooting. For this purpose, they generated a grid of models with instantaneous mixing in the overshoot region, with varying masses, initial hydrogen abundances, central hydrogen contents, and overshooting efficiencies. They showed that no model of the grid was able to reproduce the period spacings of the reference model computed with diffusive overshoot (see Fig. \ref{fig_pedersen18}). This shows that for this type of star, one should be able to distinguish between an instantaneous and a diffusive overshoot. This is no longer true for more evolved models nearing the end of the main sequence (\citealt{pedersen18}).

In principle, information could also be obtained about the temperature stratification in the overshooting region. Indeed, with penetrative convection ($\nabla = \nabla_{\rm ad}$), the \vaisala\ frequency vanishes in the overshooting region, whereas with non-penetrative overshooting ($\nabla = \nabla_{\rm rad}$), it remains strictly positive. The inner turning point $r_{\rm i}$ of the g-mode cavity is therefore located deeper in the latter case. This should have an impact on the \textit{buoyancy radius} of the sharp $\mu$-gradient, defined as $\Pi_\mu = \left( \int_{r_{\rm i}}^{r_\mu} \frac{N}{r} \,\hbox{d}r \right)^{-1}$, and thus on the period of the oscillatory behavior of $\Delta\Pi_l$, which corresponds to the ratio between the buoyancy radius of the glitch and the total buoyancy radius of the cavity. This remains to be theoretically addressed.

To use the information conveyed by $\gamma$ Doradus and SPB stars about the core properties, one difficulty arises: these stars are usually fast rotators and the effects of rotation need to be taken into account to properly identify and interpret the periodic modulation caused by the $\mu$-gradient region. This issue has been extensively studied and goes beyond the scope of the present review. However, we can mention that the validity of the so-called \textit{traditional approximation of rotation} (TAR, \citealt{eckart60})\footnote{This approximation consists in assuming a spherical shape for the star and neglecting the horizontal component of the rotation vector. This way, the problem remains separable in the radial and latitudinal coordinates, as it is for slow rotators.} has been shown (\citealt{ballot12}). This has made it possible to successfully identify the modes and analyze the oscillation spectra of fast-rotating $\gamma$ Doradus and SPB stars (\citealt{bouabid13}).

%For stars massive enough for the CNO cycle to dominate during their whole evolution, the convective core recedes during the main-sequence evolution, which increases the size of the g-mode cavity. According to Eq. \ref{eq_deltap}, this causes the period spacing of g modes to decrease during the evolution. 

\subsection{Some recent results}

\noindent \textit{HD50230:} This star is a hybrid pulsator, oscillating both as an SPB star (gravity modes) and a $\beta$ Cephei star (pressure modes) orbserved with \corot. It is also a slow rotator, which simplifies the interpretation of its oscillation spectrum. In the g-mode region of the spectrum, a group of eight modes with nearly constant period spacing was found by \cite{degroote10}. The period spacings of these modes show a periodic modulation that the authors attributed to the edge of the mixed more. The authors found that the period of this modulation can only be accounted for with extra-mixing beyond the convective core over a distance of at least $0.2\,H_P$. Interestingly, the amplitude of the modulation seems to decrease with increasing period, which the authors interpreted as an evidence for a smooth gradient of chemical composition at the boundary of the mixed core. \\

%\begin{figure}
%\begin{minipage}{6.5cm}
%\centering
%\includegraphics[width=6.5cm]{fig_degroote10.ps}
%\caption{Period spacings of eight consecutive g modes detected in the spectrum of the hybrid B pulsator HD50230 (from \citealt{degroote10}). \label{fig_degroote10}}
%\end{minipage}
%\hfill
%\begin{minipage}{9.5cm}
%\centering
%\includegraphics[width=9.5cm]{fig_morav16b.ps}
%\caption{bla (from \citealt{moravveji16}). \label{fig_morav16b}}
%\end{minipage}
%\end{figure}

\begin{figure}
\centering
\includegraphics[width=7cm]{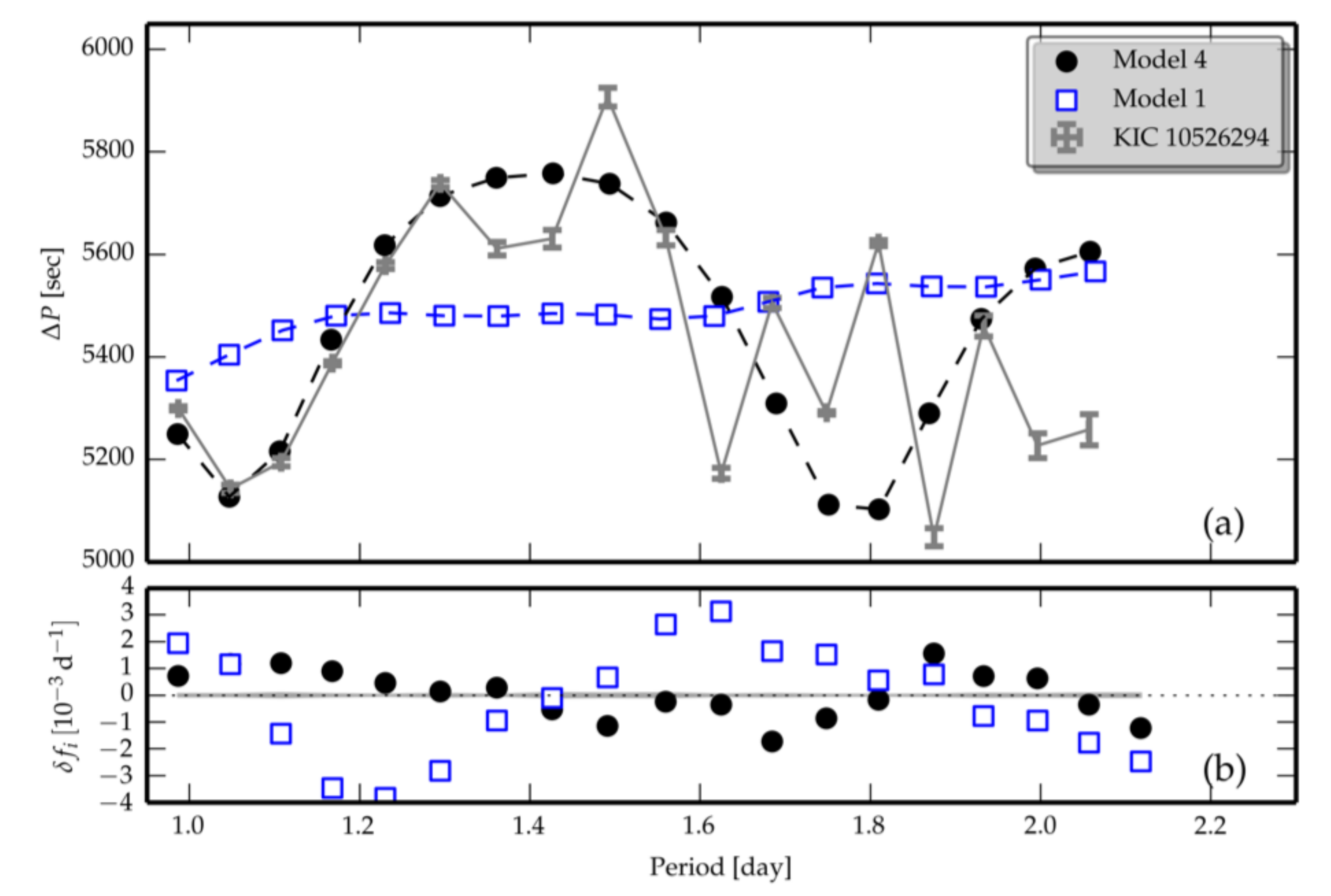}
\includegraphics[width=9cm]{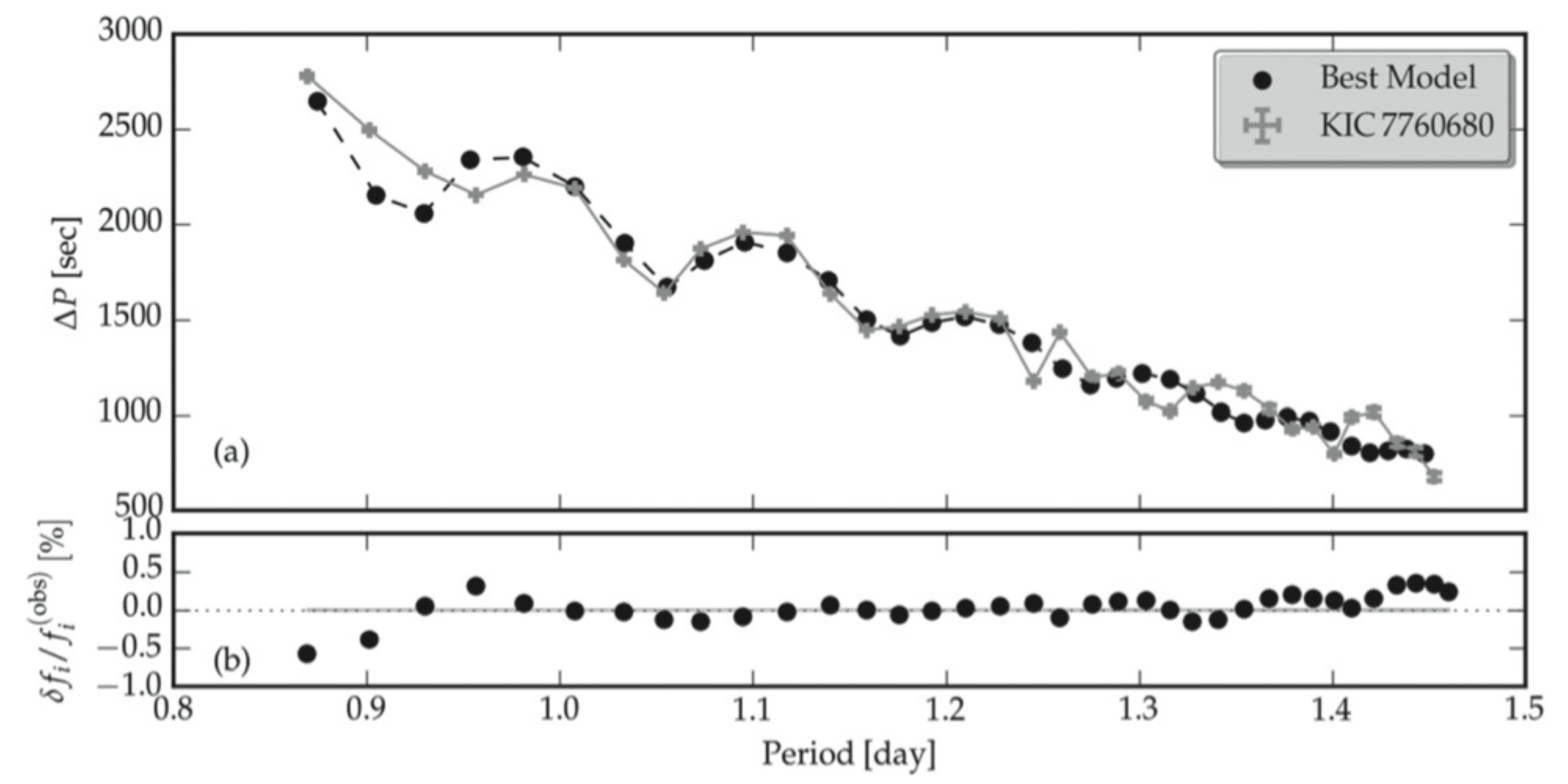}
\caption{Period spacings of consecutive dipolar g modes detected in the spectra of the SPB stars KIC10526294 (left) and KIC7760680 (right). The black circles (resp. blue squares) show the mode periods for the best model obtained with (resp. without) diffusive overshooting. The bottom plots show the residuals. Figures from \cite{moravveji15} (left) and \cite{moravveji16} (right). 
\label{fig_morav}}
\end{figure}

\noindent \textit{KIC10526294:} KIC10526294 is an SPB star observed with \kepler. A series of 19 dipolar gravity modes with consecutive radial orders were detected by \cite{papics14} for this star, making it a particularly interesting target to search for periodic modulation induced by the convective core. Rotational splittings could be measured for the star, which indicated that it is a very-slow rotator (average rotation period of $\sim$ 188 days). The period spacings $\Delta P$ of the detected modes exhibit a clear deviation from the asymptotic period spacing. \cite{moravveji15} performed a detailed modeling of this target. They showed that the variations of $\Delta P$ with mode period are better reproduced with core overshooting implemented as a diffusive process than with an instantaneous mixing in the overshoot region. They found optimal values of the overshoot parameters of $\fov$ between 0.017 and 0.018 (see Eq. \ref{eq_ov_exp}). They also claim that including an extra-mixing in the radiative interior outside the overshooting region can significantly improve the agreement between the models and the observations. It should however be remarked that the optimal models are still far from giving a good statistical agreement with the \kepler\ observations (see Fig. \ref{fig_morav}, left panel). This suggests that the models might be missing some important ingredient. \\

\noindent \textit{KIC7760680:} This star is a moderately-rotating SPB star observed with the \kepler\ satellite. It exhibits a series of 36 consecutive gravity modes, in which a clear periodic modulation can be detected (see Fig. \ref{fig_morav}, right panel). It is also apparent that the period spacings of KIC7760680 show an almost linear decrease with mode period. This is the clear signature of moderate rotation for prograde modes (\citealt{bouabid13}). \cite{moravveji16} modeled the star, considering different assumptions for the mixing within the overshooting region. They considered a solid-body rotation for the star and for each model, they optimized the rotation rate to reproduce the slope of the period spacings as a function of the mode period. As was the case for HD50230 and KIC10526294, they found that a diffusive overshoot reproduces the periodic modulation in the period spacings better than an instantaneous overshoot. With both implementations, the optimal models include a sizable overshooting region ($\fov = 0.024\pm0.001$ in the case of a diffusive overshoot and $\aov\sim0.32$ for an instantaneous overshoot). Here again, the optimal solutions are quite far from the observations, yielding reduced $\chi^2$ of the order of 2000. The bottom right panel of Fig. \ref{fig_morav} shows that there is clear structure in the residuals (periodic modulation for mode periods larger than $\sim$1.25 days). This shows that the period of the modulation in $\Delta P$ differs between the models and the observations, especially for large mode periods. This is likely indicating that improvements could be made in the modeling of the chemical composition profile in the overshooting region. \\

\noindent \textit{$\gamma$ Doradus stars:} Recently, long series of consecutive g modes were also revealed in the spectra of $\gamma$ Doradus stars (\citealt{vanreeth16}, \citealt{christophe18}). These stars are generally moderate to fast rotators. However, once the signature of rotation has been correctly identified, an oscillatory behavior of the period spacings has been reported for some $\gamma$ Doradus stars (\citealt{christophe18}). These stars could therefore also provide precious information on the properties of the extended mixed cores in the near future.

%\subsection{Constraints from $\beta$ Cephei stars \label{sect_betacep}}
%
%$\beta$ Cep: previous constraints, HD129929 \cite{aerts03}, \cite{dupret04}, $\nu$ Eri \cite{ausseloos04}

\section{Subgiants \label{sect_subgiants}}

When stars evolve past the end of the main sequence, their inner layers contract as hydrogen starts burning in a shell. This causes the frequencies of gravity modes to increase owing to the increasing \vaisala\ frequency in the core. In the meantime, the envelope expends as stars become subgiants. The mean density of the star decreases and therefore the frequencies of pressure modes also decrease. As a result, the frequencies of the lowest radial order g modes become of the same order of magnitude as the frequencies of the p modes that are stochastically excited in the outer part of the convective envelope. At this point, non-radial modes develop a \textit{mixed nature}, behaving as g modes in the core and as p modes in the envelope. This phenomenon arises because of the coupling exerted between the two cavities by the evanescent zone that separates them. Mixed modes have a large potential because they convey information about the core properties while having detectable amplitudes at the surface.

\subsection{What constraints can we expect from mixed modes?}

The helium core of subgiants is radiative because hardly produces any luminosity. So even if the star had a convective core during the main sequence, convective mixing has ceased when the star becomes a subgiant. Nevertheless, the main sequence convective core leaves an imprint in the chemical composition profile of young subgiants. Since mixed modes are sensitive to the \vaisala\ profile, and thus to the profile of $\mu$, they can bring indirect information about the extent of the core and the nature of the mixing at its edge.

\begin{figure}
\begin{minipage}{8cm}
\centering
\includegraphics[width=8cm]{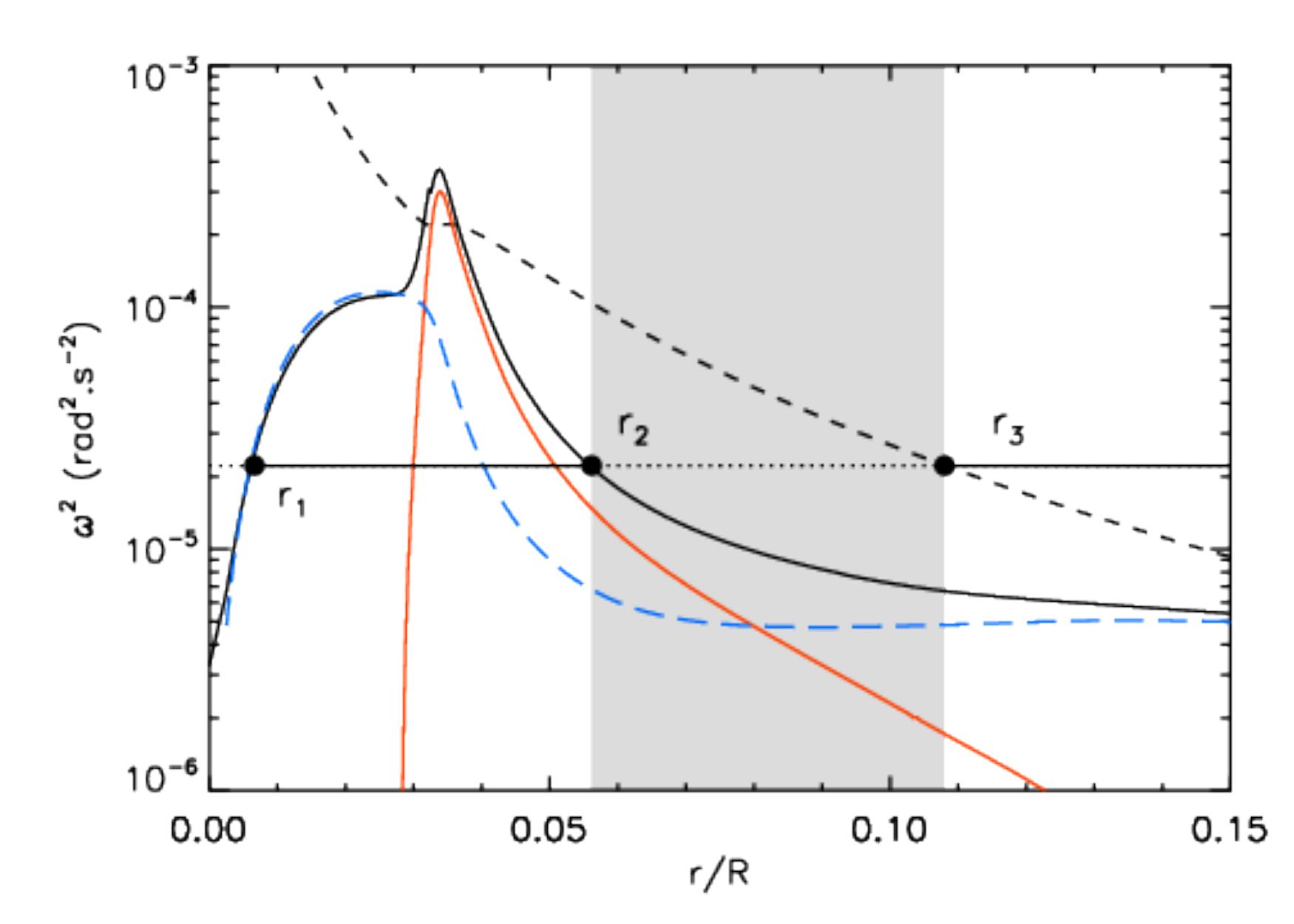}
\caption{Propagation diagram of a $1.3\,M_\odot$ subgiant. The \vaisala\ frequency (black curve) is split into its thermal part (blue dashed line) and its chemical part (red solid line). The $l=1$ Lamb frequency is shown by the black dashed line. The horizontal line indicates the frequency of an $l=1$ mixed mode with dotted lines showing evanescent regions. Figure from \cite{deheuvels11}. \label{fig_deheuvels11a}}
\end{minipage}
\hfill
\begin{minipage}{8cm}
\centering
\includegraphics[width=8cm]{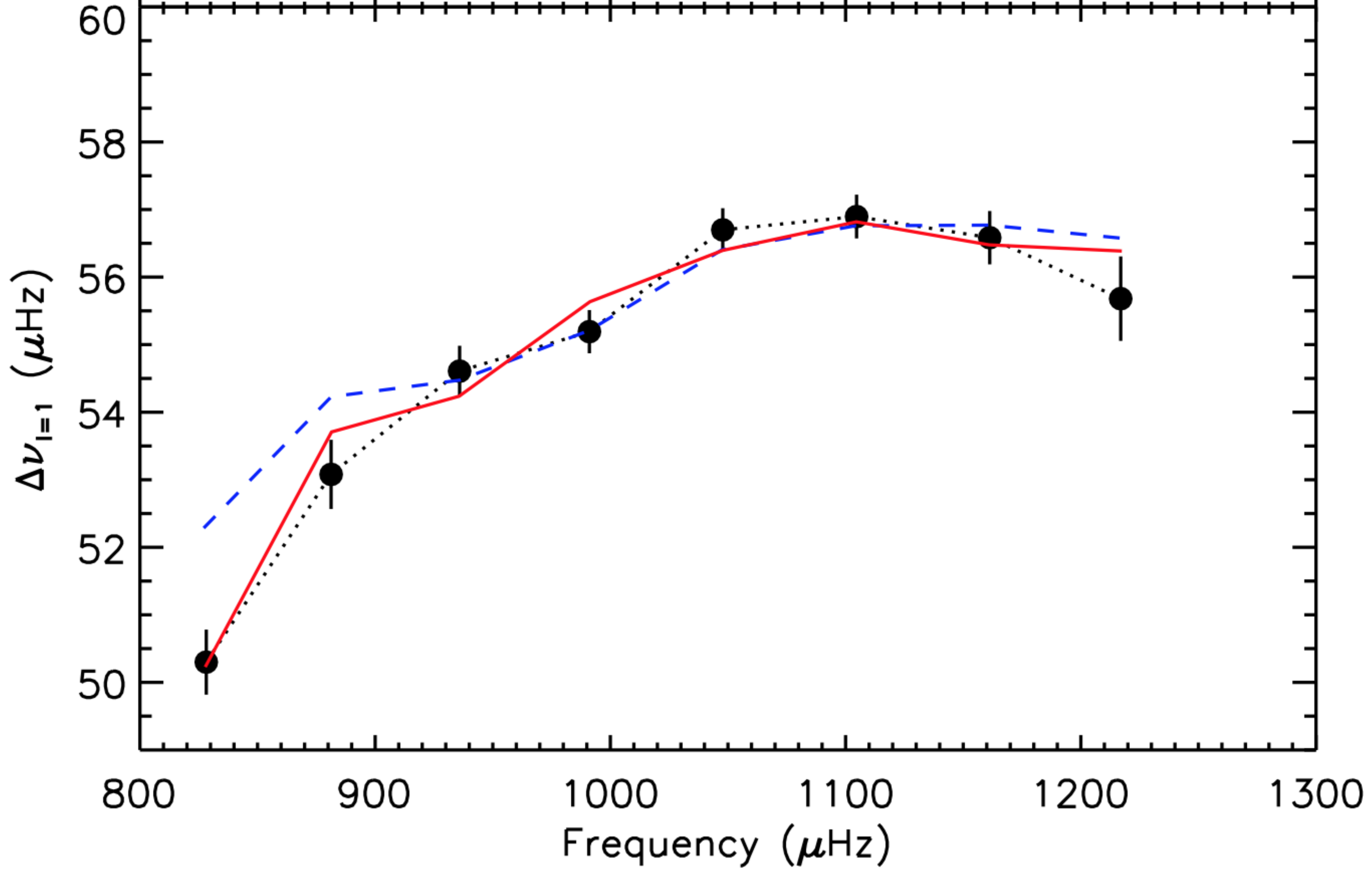}
\caption{Variations in the large separation of $l=1$ modes as a function of mode frequency for HD49385. The black dots correspond to \corot\ data, the red solid line shows $\Delta\nu_1$ for the best-fit model, obtained with $\aov=0.19$, and the blue dashed line corresponds to the best model with $\aov=0.1$. Figure from \cite{deheuvels11}. \label{fig_deheuvels11b}}
\end{minipage}
\end{figure}

The oscillation spectra of young subgiants contain only a few g-dominated modes, i.e., modes that are trapped mainly in the g-mode cavity. However, in subgiants, the coupling between the p- and g-mode cavities is strong for dipolar modes, and the frequencies of p-dominated modes are significantly affected by this coupling (\citealt{rome}). Mixed modes convey information about the core properties through two channels:
\begin{itemize}
\item \textit{The frequencies of g-dominated modes.} As is apparent from Eq. \ref{eq_deltap}, they depend essentially on the integral $\int_{r_1}^{r_2} N/r \,\hbox{d}r$, where $r_1$ and $r_2$ are the inner and outer turing points of the g-mode cavity. Fig. \ref{fig_deheuvels11a} shows the \vaisala\ profile of a $1.3\,M_\odot$ model in the subgiant phase. In the outer part of the g-mode cavity (below $r_2$), the \vaisala\ frequency is dominated by the contribution of the $\mu$-gradient ($N^2_\mu = g\varphi\nabla_\mu/H_P$, red solid line), whose shape depends on the extent of the main sequence convective core.
\item \textit{The intensity of the coupling between the p- and g-mode cavities.} The coupling essentially depends on the \vaisala\ profile in the evanescent zone ($r_2\leqslant r \leqslant r_3$ in Fig. \ref{fig_deheuvels11a}). It thus conveys information about the $\mu$-gradient above $r_2$, as can be seen in Fig. \ref{fig_deheuvels11a}. The intensity of the coupling can be estimated observationally by observing its effect on the p-dominated modes. For low coupling intensities, their frequencies will hardly deviate from the asymptotic frequencies of p modes, whereas if the coupling is strong, large deviations are expected.
\end{itemize}

\subsection{Recent results}

The star HD49385 was observed with the \corot\ satellite during 137 days and its oscillation spectrum was analyzed by \cite{analyse_49385}. Fig. \ref{fig_deheuvels11b} shows the variations in the large separation $\Delta\nu_1$ of dipolar modes as a function of mode frequency. At low frequency, $\Delta\nu_1$ strongly deviates from the roughly constant value that is expected from asymptotic developments. It was later established that this was caused by the presence of a g-dominated mixed mode in the lower-frequency part of the spectrum, which coupled to the detected p modes and altered their mode frequencies (\citealt{rome}).

\cite{deheuvels11} proposed a new optimization technique adapted to the modeling of stars with mixed modes, which they applied to HD49385. They found that the star has a mass of $1.25\pm0.05\,M_\odot$ and an age of $5.0\pm0.3$ Gyr. For their modeling, the authors considered models with an instantaneous overshooting over an adjustable distance $\dov$. They found two different families of solutions: one with a small amount of overshooting ($\aov<0.05$) and the other with a moderate amount of overshooting ($\aov=0.19\pm0.01$). The models from the latter family provide the closest agreement with the observations and the large separation of their $l=1$ modes are shown in Fig. \ref{fig_deheuvels11b}. \cite{deheuvels11} showed that this bimodality of the solutions is due to the strong dependence of the mode coupling to the stellar mass (the higher the mass, the lower the coupling). Only models with masses around $1.25\,M_\odot$ are able to produce the correct coupling and thus reproduce the observed frequencies of $l=1$ modes. The optimal mass was found to vary non-linearly with the amount of overshooting. Only low ($\aov<0.05$) or moderate ($\aov=0.19\pm0.01$) values of overshooting correspond to a stellar mass of about $1.25\,M_\odot$. Models with $\aov\sim0.1$ have higher masses and thus a mode coupling that is too weak (see blue dashed curve in Fig. \ref{fig_deheuvels11b}). Models with $\aov>0.2$ have lower masses and thus a coupling that is too strong.

Mixed modes can thus give measurements of the size of main sequence convective cores using a diagnostic that is completely independent from the one used for main sequence solar-like pulsators (Sect. \ref{sect_solarlike}). Several tens of subgiants have been observed with \kepler\ and could also provide constraints on the size of main sequence convective cores. The study of these targets is under way.

\section{Core-He burning giants \label{sect_CHeB}}

Giant stars with masses $M\gtrsim0.7 M_\odot$ eventually start burning helium in their core. This happens either quietly in a non-degenerate core (for stars with masses $M\gtrsim 2 M_\odot$) or in a flash for stars with masses $M\lesssim 2 M_\odot$, whose core is degenerate when it reaches the temperature at which He starts burning. In both cases, the star then develops a convective core. Measuring the extent of the mixed core at this evolutionary stage can bring complementary information about the interface between convective and radiative regions.
%whose boundary can be extended owing to the same mechanisms that operate during the main sequence. 
We start by briefly introducing the challenges posed by the modeling of convective cores in core-helium burning (CHeB) stars (Sect. \ref{sect_core_clump}) and we then present the constraints derived from asteroseismology (\ref{sect_clump_sismo}).

\subsection{Modeling the convective core of CHeB giants \label{sect_core_clump}}

\begin{figure}
\centering
\includegraphics[width=10cm]{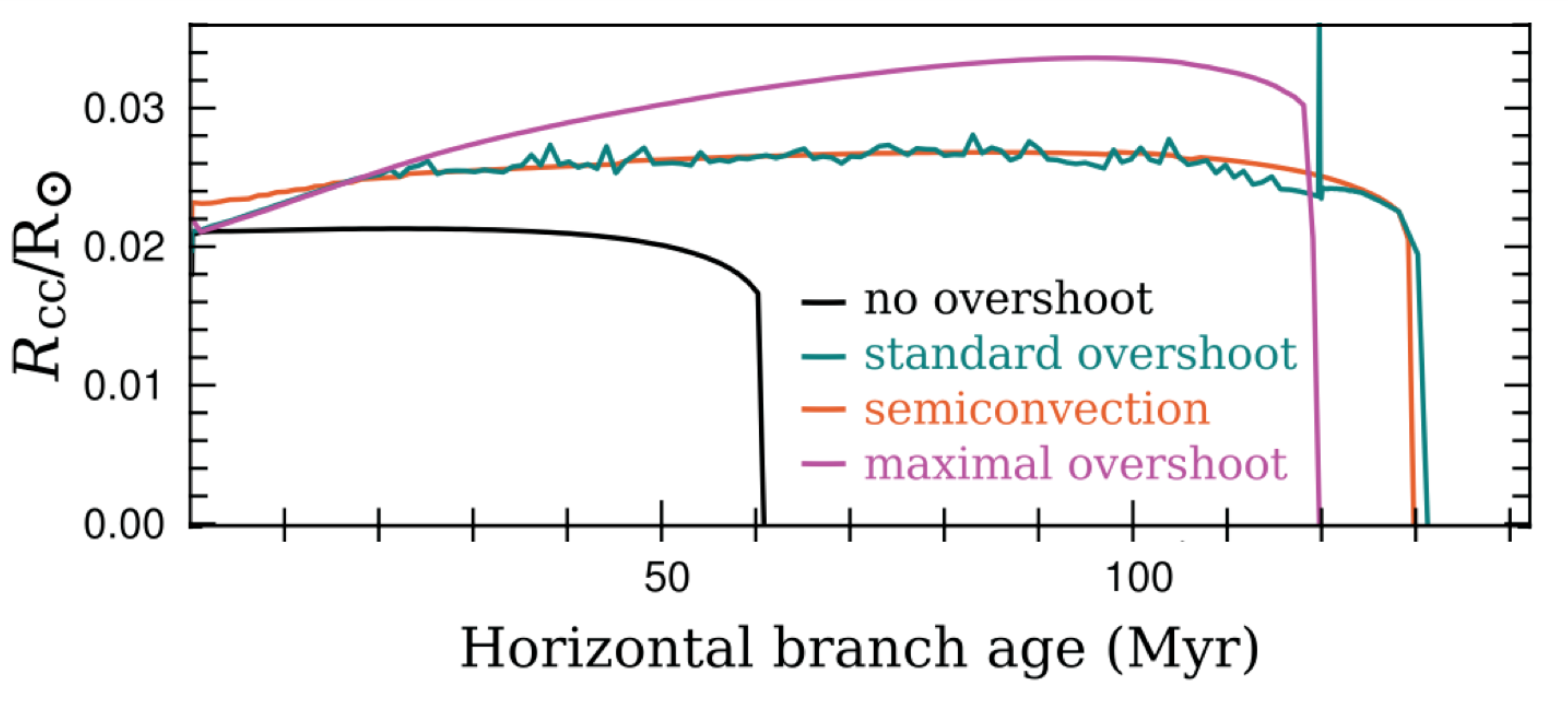}
\caption{Size of the mixed core during the CHeB phase with different modelings for the boundary of the convective core. Figure from \cite{constantino15}. 
\label{fig_constantino15}}
\end{figure}

The modeling of mixing in the core of low- and intermediate-mass stars during the CHeB phase is notoriously challenging. Depending on the criterion that is adopted for convective stability, evolutionary codes predict very different values for the size of the He-burning convective core, and thus also for the duration of the CHeB phase (see Fig. \ref{fig_constantino15}). The situation is more complicated than during the main sequence because C and O, which accumulate as He is burnt in the core, are more opaque than He. As a result, the radiative gradient increases in the convective core, and a discontinuity of the radiative gradient tends to develop at the boundary of the convective core. We here briefly describe some of the choices made to treat this in evolutionary codes and refer the interested reader to the review by \citealt{salaris17} for more details.

\begin{figure}
\begin{minipage}{8cm}
\centering
\includegraphics[width=7cm]{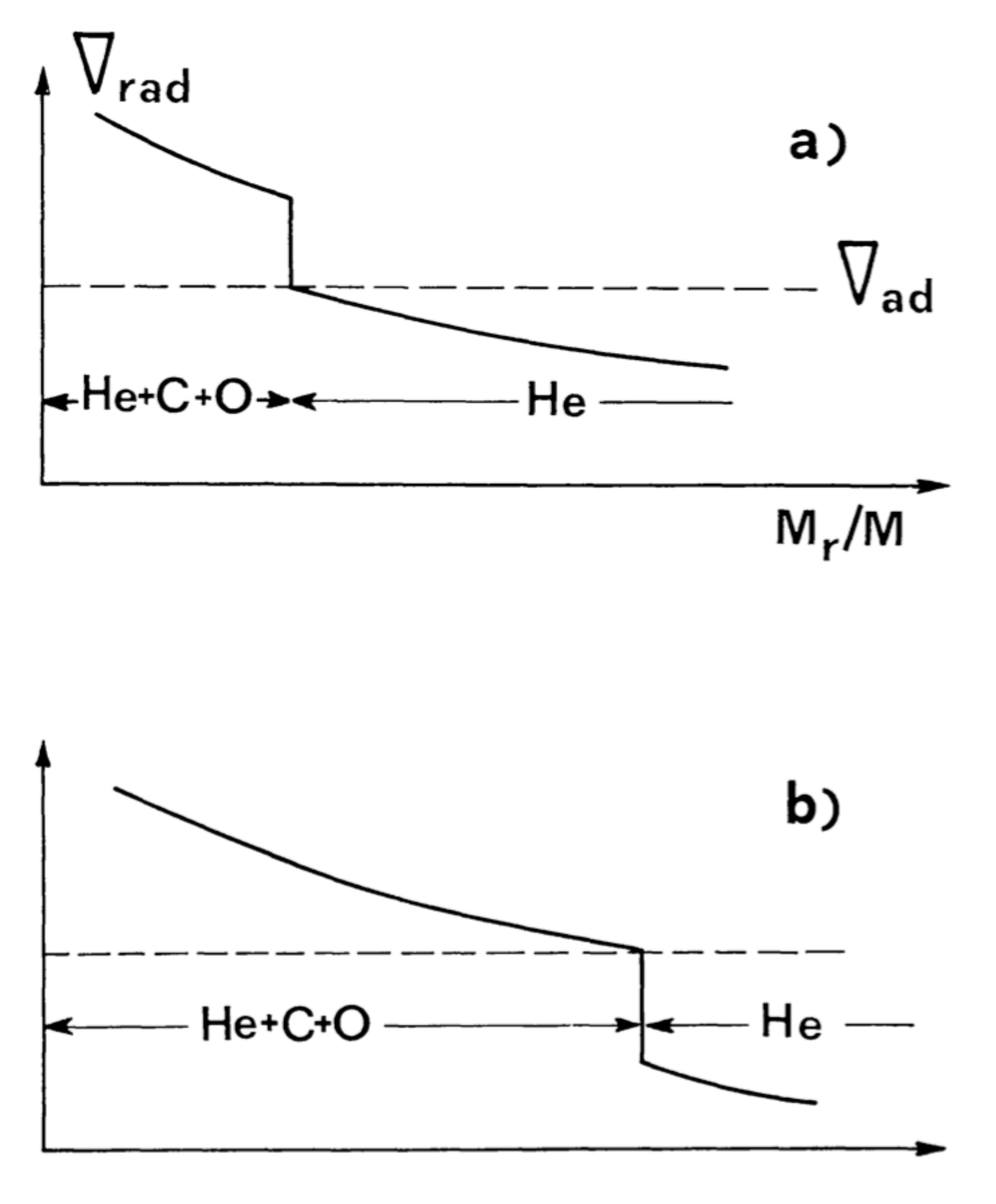}
\caption{Schematic behavior of the temperature gradient near the boundary of the convective core with $\nabla_{\rm rad}=\nabla_{\rm ad}$ imposed on the radiative side (a) and on the convective side (b) of the boundary (from \citealt{castellani71a}). \label{fig_castellani71a}}
\end{minipage}
\hfill
\begin{minipage}{8cm}
\centering
\includegraphics[width=6cm]{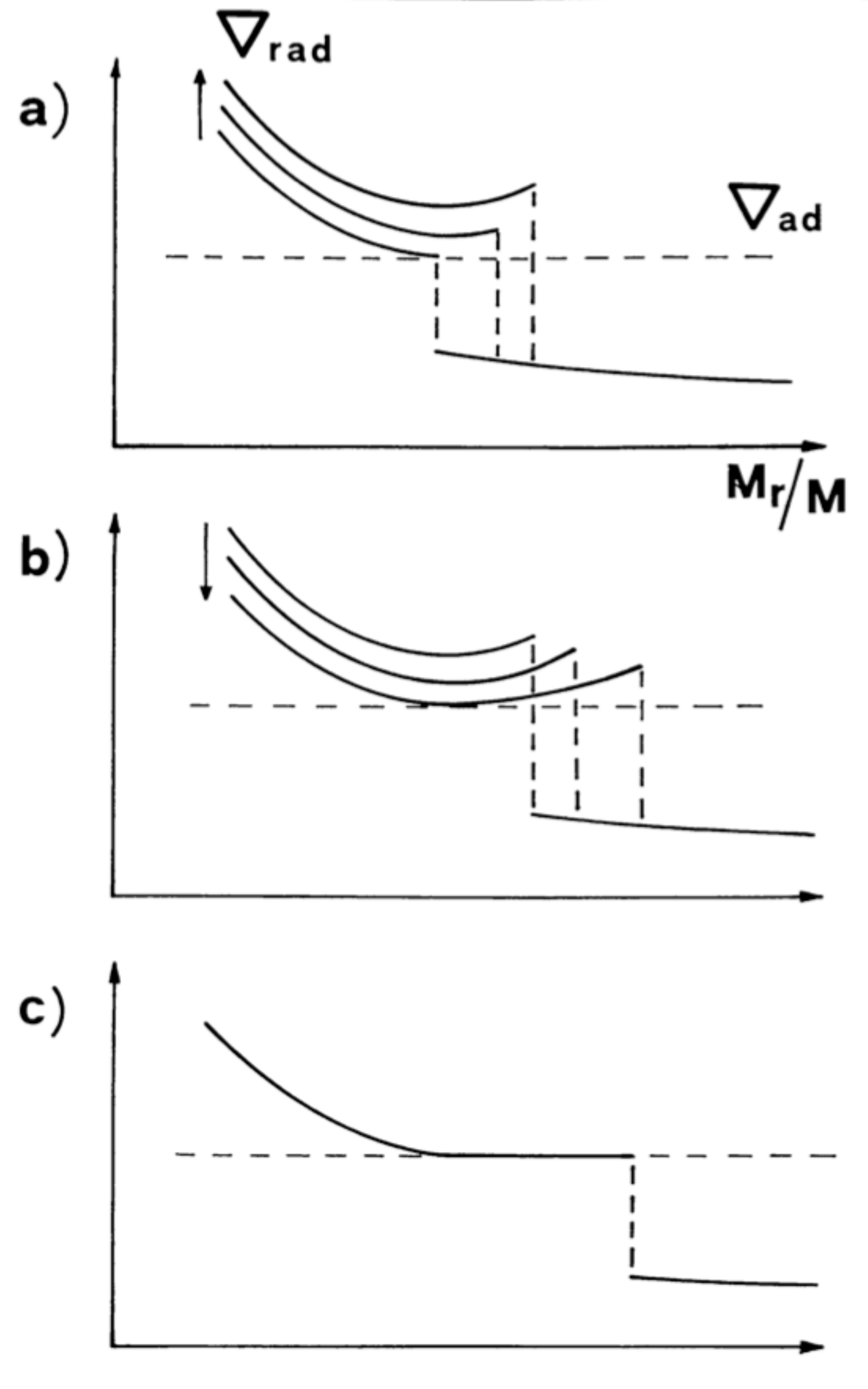}
\caption{Schematic behavior of $\nabla_{\rm rad}$ after it has reached a minimum in the convective core. Panel a (reps. b) shows the evolution with standard overshooting and an increasing (resp. decreasing) radiative gradient. Panel c: evolution with semiconvection. Figure from \cite{castellani71b}. \label{fig_castellani71b}}
\end{minipage}
\end{figure}

In what is usually referred to as the \textit{bare Schwarzschild} (BS) model, the Schwarzschild criterion is applied \textit{on the radiative side} of the convective boundary (panel (a) of Fig. \ref{fig_castellani71a}). Since the radiative gradient is hardly modified over time on the radiative side, the core size remains roughly constant during the whole CHeB phase (see black curve in Fig. \ref{fig_constantino15}, labeled as the ``no ivershooting'' case). Meanwhile, the radiative gradient increases in the core, and the quantity $\nabla_{\rm rad} - \nabla_{\rm ad}$ thus increases \textit{on the convective side} of the core boundary. As established by \cite{schwarzschild58} and reminded by \cite{castellani71a} and \cite{gabriel14}, this situation is in fact unphysical because the convective velocities are expected to vanish at the edge of the convective core. As a result, the total flux should be equal to the radiative flux at this layer, and one should have $\nabla_{\rm rad} = \nabla \approx \nabla_{\rm ad}$ there. The BS model is therefore an incorrect implementation of the Schwarzschild criterion.

Another way of understanding the inadequacy of the BS model is to realize that it is unstable to any mixing beyond the core boundary. Indeed, let us assume a mild extra-mixing, such that the first layer above the convective core is mixed with the convective core. In this layer, the abundance in carbon and oxygen increases, the opacity increases and hence the radiative gradient increases above the adiabatic gradient. The layer then becomes definitively convective. At the next time step, the layer above the enlarged convective core will in turn become convective. This process stops only when $\nabla_{\rm rad}=\nabla_{\rm ad}$ \textit{on the convective side} of the core boundary. Panel (b) of Fig. \ref{fig_castellani71a} thus shows the correct implementation of the Schwarzschild criterion. In practice, this is implemented in evolution codes by including a small amount of core overshooting (the extension of the convective core that it produces is sometimes referred to as \textit{induced overshooting}) or by checking at each time step whether the layers above the convective core would become convective if they were mixed with the core, and by adding these layers to the core if it is the case.

%\begin{figure}
%\centering
%\includegraphics[width=10cm]{fig_bossini15.ps}
%\caption{Ratio between $\nabla_{\rm rad}$ and $\nabla_{\rm ad}$ for different models along the evolution during the CHeB phase. The equality $\nabla_{\rm rad}=\nabla_{\rm ad}$ is imposed either on the radiative side (bare Schwarzschild model, left panel) or on the convective side (right panel). Figure from \cite{bossini15}. 
%\label{fig_bossini15}}
%\end{figure}

However, a complication occurs when the mass fraction of helium in the core drops below $\sim 0.7$. Then, a minimum appears in the profile of the radiative gradient in the core, as can be seen in panel (a) of Fig. \ref{fig_castellani71b}. For low and intermediate amounts of overshooting, the outward mixing brings fresh helium into the core and thus induces a decrease of $\nabla_{\rm rad}$ in the whole convective core (panel (b) of Fig. \ref{fig_castellani71b}). The minimum of $\nabla_{\rm rad}$ eventually drops below $\nabla_{\rm ad}$ and the convective core is split in two convective regions separated by an intermediate radiative zone. The outer convective region rapidly vanishes because of the decrease in $\nabla_{\rm rad}$. The convective core is thus comprised only of the inner convective region and it has shrunk. As helium is burnt in the core, the abundance of carbon and oxygen increases again, $\nabla_{\rm rad}$ increases and eventually has again a minimum within the core. We are then brought back to panel (a) of Fig. \ref{fig_castellani71b} and the situation reiterates. As a result, the boundary of the convective core goes back and forth (see the case labeled as \textit{standard overshooting} in Fig. \ref{fig_constantino15}), leaving behind step-like features in the helium abundance profile. In this case, the behavior of the convective core is in fact independent of the amount of core overshooting that is included (this is no longer true for large amounts of overshooting as explained below).

The treatment of the intermediate radiative region that appears in the vicinity of the minimum of $\nabla_{\rm rad}$ has been the subject of several studies. It is generally thought that it undergoes a partial mixing that enforces convective neutrality ($\nabla_{\rm rad} = \nabla_{\rm ad}$) in this zone (\citealt{castellani71b}, \citealt{castellani85}), as can be seen in panel (c) of Fig. \ref{fig_castellani71b}. The partially mixed region shares similar features with a semi-convective layer, and this mechanism has been referred to as \textit{induced semi-convection}. Modeling this intermediate region as a semi-convective layer produces core sizes that are very similar to those obtained with overshooting (see how the cyan and orange curves nearly overlap in Fig. \ref{fig_constantino15}), but without the back-and-forth motion of the core boundary, and therefore with a smoother chemical composition profile.

It was also found that when applying large amounts of core overshooting at the boundary of the convective core, the extra-mixed region becomes large enough to prevent the formation of a semi-convective region (\citealt{bressan86}, \citealt{bossini15}). In this case, the size of the mixed core depends on the amount of core overshooting that is imposed.

\subsection{Constraints from asteroseismology \label{sect_clump_sismo}}

\subsubsection{Asymptotic period spacings of g modes in CHeB giants}

Thanks to the space missions \corot\ and \kepler, mixed modes have now been detected in tens of thousands of red giants. The frequencies of these modes can be identified using their asymptotic expression, which was first developed by \cite{shibahashi79}. By fitting this analytic expression to the observed mode frequencies, one can obtain estimates of various global seismic characteristics of the star, including the asymptotic period spacing $\Delta\Pi_1$ of its dipolar gravity modes (see Eq. \ref{eq_deltap}). The fitting procedure is challenging because of the large number of modes and it is made much more complicated by the splitting of mixed modes due to rotation. \cite{mosser15} have proposed a convenient method, based on the calculation of corrected mode periods (called \textit{stretched periods}), which made it possible to perform an automatic fitting of red giants. Using this method, \cite{vrard16} were able to measure the asymptotic period spacing $\Delta\Pi_1$ of 6100 \kepler\ red giants.

This database constitutes an unprecedented opportunity to probe the core of red giants. \cite{bedding11} showed that the period spacing $\Delta\Pi_1$ can be used to reliably distinguish CHeB giants from H-shell burning giants, which are ascending the red giant branch (RGB). The reason for this is evident from Eq. \ref{eq_deltap}. In contrast with RGB stars, CHeB giants have a convective core. Their g-mode cavity is therefore smaller and they have larger values of $\Delta\Pi_1$. For CHeB giants, \cite{montalban13} showed that there is a nearly linear relation between the size of the convective core and the asymptotic period spacing $\Delta\Pi_1$. Indeed, if the convective core expends, the g-mode cavity becomes smaller and $\Delta\Pi_1$ increases. The \kepler\ data thus have a great potential to measure the size of the mixed core in CHeB stars.

The asymptotic period spacings can also convey information about the temperature stratification. Indeed, in the case of penetrative convection, we have $\nabla=\nabla_{\rm ad}$ and thus $N^2 = 0$ in the extra-mixed region. As a result, gravity waves do not propagate in the overshoot region. On the contrary, with non-penetrative overshooting, $N^2 = N_T^2>0$ and the overshoot region is part of the g-mode cavity. We thus expect models with non-penetrative convection to have smaller values of $\Delta\Pi_1$ than  models computed with penetrative convection. For models with semi-convection above the convective core, $N^2 = N_\mu^2 >0$ in the partially mixed region and $\Delta\Pi_1$ is also expected to be smaller than with penetrative overshooting.

As described in Sect. \ref{sect_gmodes}, sharp variations in the \vaisala\ frequency (buoyancy glitches) induce periodic modulations in the period spacings of g modes. Such features could be measured from the frequencies of mixed modes and give strong constraints on the chemical composition profile near the core boundary. We come back to this in more detail in Sect. \ref{sect_glitch_clump}.

\subsubsection{Seismic constraints on the convective core of CHeB giants}

\begin{figure}
\centering
\includegraphics[width=8cm]{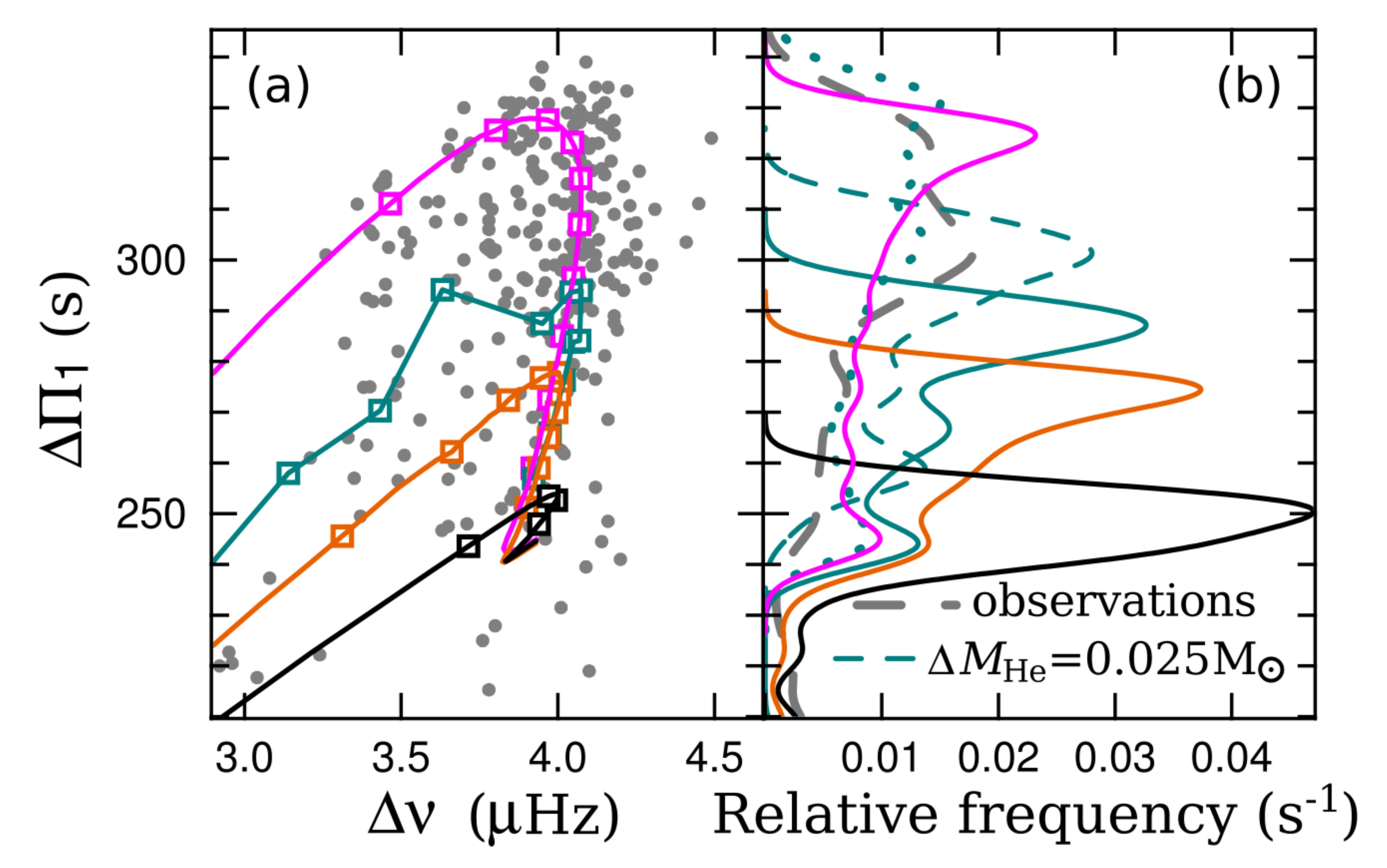}
\includegraphics[width=8cm]{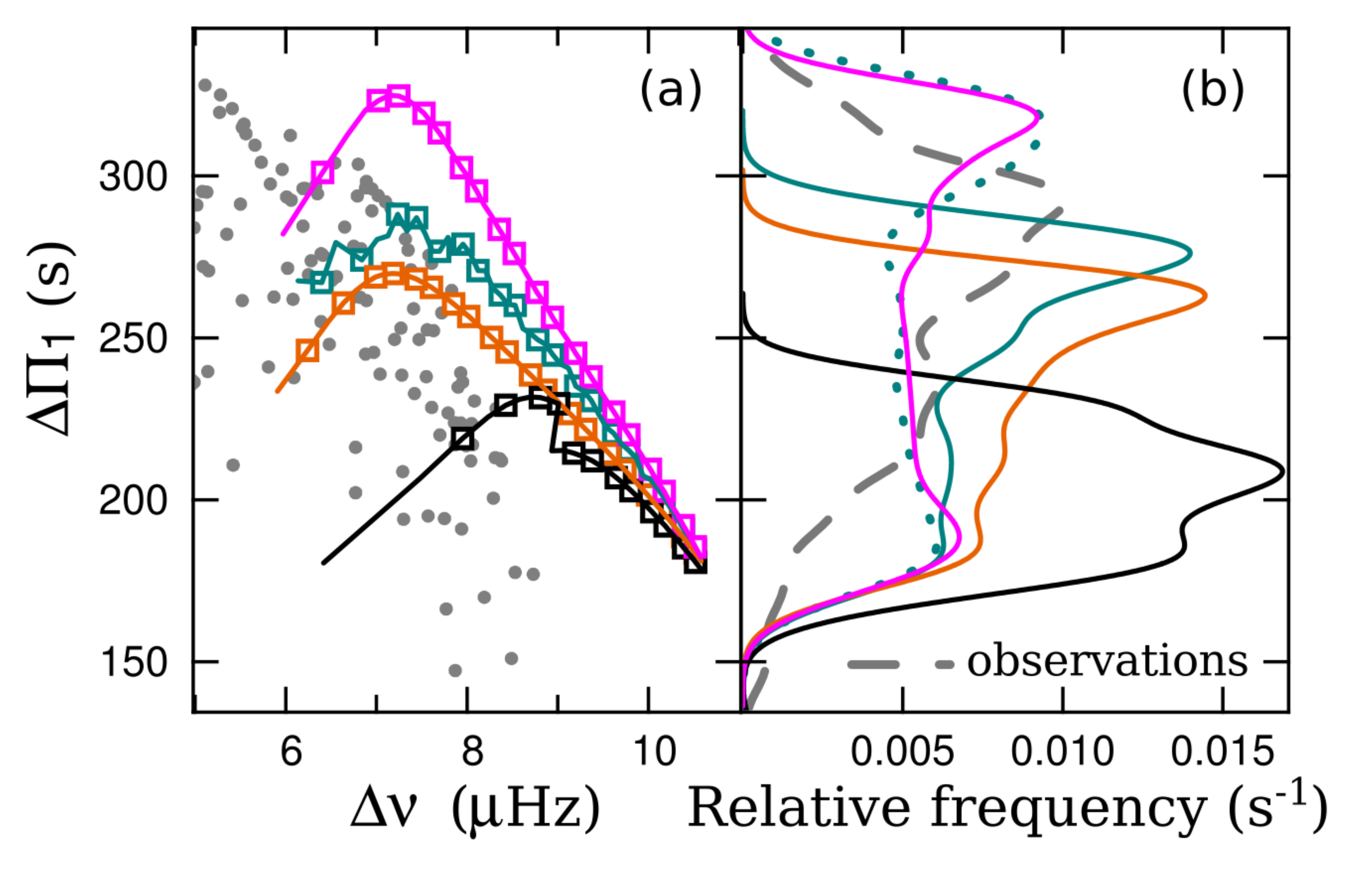}
\caption{Evolution in the $\Delta\nu$-$\Delta\Pi_1$ plane of 1\,$M_\odot$ (left figure) and 2.5\,$M_\odot$ (right figure) CHeB models computed with different mixing schemes (``bare Schwarzschild'' in black, low overshoot in cyan, semi-convection in orange, maximal overshoot in magenta). The grey dots correspond to the observed period spacings of \kepler\ giants restricted to stars that have undergone a He flash ($M\leqslant 2M_\odot$) in the left figure, and stars that triggered He-burning quietly ($M\geqslant 2M_\odot$). Panels (b) of both figures show probability density curves with the same color code (observations in thick grey dashes). Figures from \cite{constantino15}. 
\label{fig_constantino15b}}
\end{figure}

\cite{constantino15} and \cite{bossini15} both led studies to compare the observed distribution of period spacings of CHeB giants to the distributions that would be predicted with different mixing schemes beyond the convective core. They found generally consistent results. 

The ``bare Schwarzschild'' models have the smallest convective cores because the (incorrect) implementation of the Schwarzschild criterion on the radiative side prevents the core from growing. The highest period spacings $\Delta\Pi_1$ predicted by these models are around 250\,s (see black symbols in Fig. \ref{fig_constantino15b}), well below the maximum observed period spacings, which are around 340 s. \cite{bossini15} reach the same conclusion. This confirms that the convective cores of the bare Schwarzschild models are much too small.

Models that include low amounts of core overshooting or semi-convection also have period spacings that appear to be too small compared to the observations (cyan and orange symbols in Fig. \ref{fig_constantino15b}). This means that their convective cores are too small. We already mentioned in Sect. \ref{sect_core_clump} that models computed with semi-convection and models computed with low overshooting have very similar core sizes (Fig. \ref{fig_constantino15}). Yet Fig. \ref{fig_constantino15b} shows that the latter models have larger period spacings. According to \cite{constantino15}, this is justified by the fact that large $\mu$-gradients develop in models computed with overshooting, owing to the back-and-forth motion of the core boundary. This is enough to create efficient mode trapping inside the partially mixed region. As a result, the observed period spacing corresponds to the asymptotic expression of Eq. \ref{eq_deltap} calculated excluding the region of $\mu$-gradient. It is thus larger than for models computed with low overshooting than for models computed with semi-convection, for which the chemical composition profile is smooth and such mode trapping does not occur.

The \kepler\ data clearly point in favor of an extended mixed core, larger than the one produced with semi-convection or standard amounts of overshooting. To reproduce the seismic data, \cite{bossini15} calculated models with high amounts of overshooting. They found that models with non-penetrative convection over a distance of 1\,$H_P$ or with penetrative convection over a distance of 0.5\,$H_P$ could roughly reproduce the distribution of the observed period spacings. They gave their preference to the latter models because they also match the luminosity of the asymptotic-giant-branch (AGB) bump, which can be measured from \kepler\ data. \cite{constantino15} calculated models with a modified implementation of core overshooting. They prevented at all time the splitting of the convective core that occurs because of the minimum in $\nabla_{\rm rad}$. This model, which they refer to as \textit{maximal overshooting} has no physical justification but aims at building convective core with maximal sizes. The authors found that these models produce period spacings that are consistent with the bulk of the low-mass observations (see magenta line in Fig. \ref{fig_constantino15b}). 

\begin{figure}
\centering
\includegraphics[width=10cm]{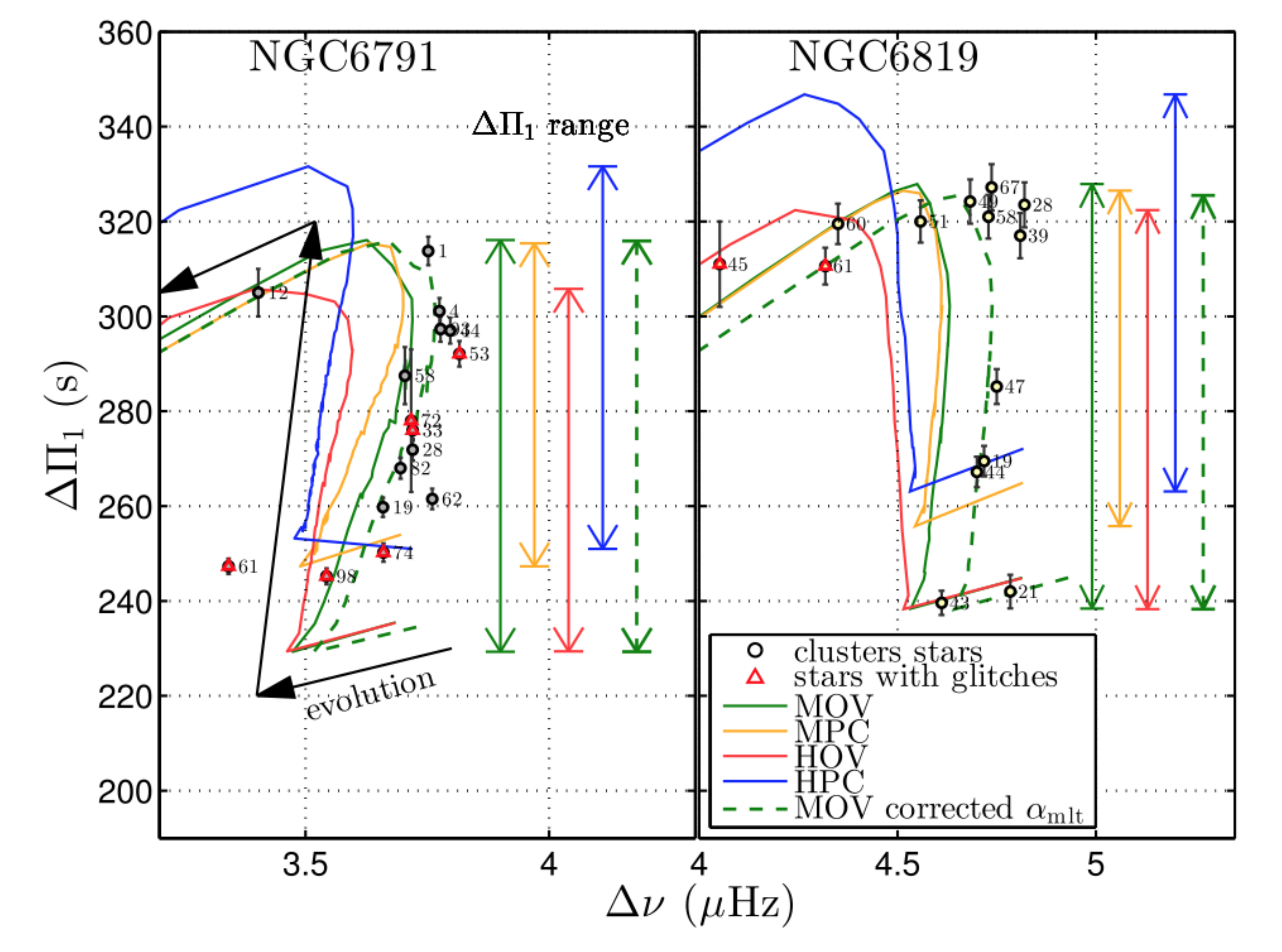}
\caption{CHeB giants of NGC\,6791 and NGC\,6819 shown in the $\Delta\nu$-$\Delta\Pi_1$ plane. Predictions from models with different mixing schemes are overplotted: intermediate (MOV) and high (HOV) non-penetrative convection, intermediate (MPC) and high (HPC) penetrative convection. Figure from \cite{bossini17}. 
\label{fig_bossini17}}
\end{figure}

Additional information was recently obtained from the measurement of period spacings in the CHeB giants of the two old open clusters NGC\,6791 and NGC\,6819 (\citealt{bossini17}). Fig. \ref{fig_bossini17} shows the location in the $\Delta\nu$-$\Delta\Pi_1$ plane of the CHeB-members of these two clusters. The authors calculated models with the same physical properties as the CHeB giants of both clusters and using different mixing schemes at the core boundary. They found that models computed with a moderate amount of overshooting can reproduce the range of observed period spacings. Interestingly, the models computed with penetrative convection (adiabatic stratification in the extra-mixed region) predict too large period spacings for the stars at the beginning of the CHeB phase in NGC\,6819, which led the authors to favor the non-penetrative convection scenario. Naturally more evidence is required to be more conclusive.

\subsubsection{Constraints from buoyancy glitches \label{sect_glitch_clump}}

\begin{figure}
\centering
\includegraphics[width=5cm]{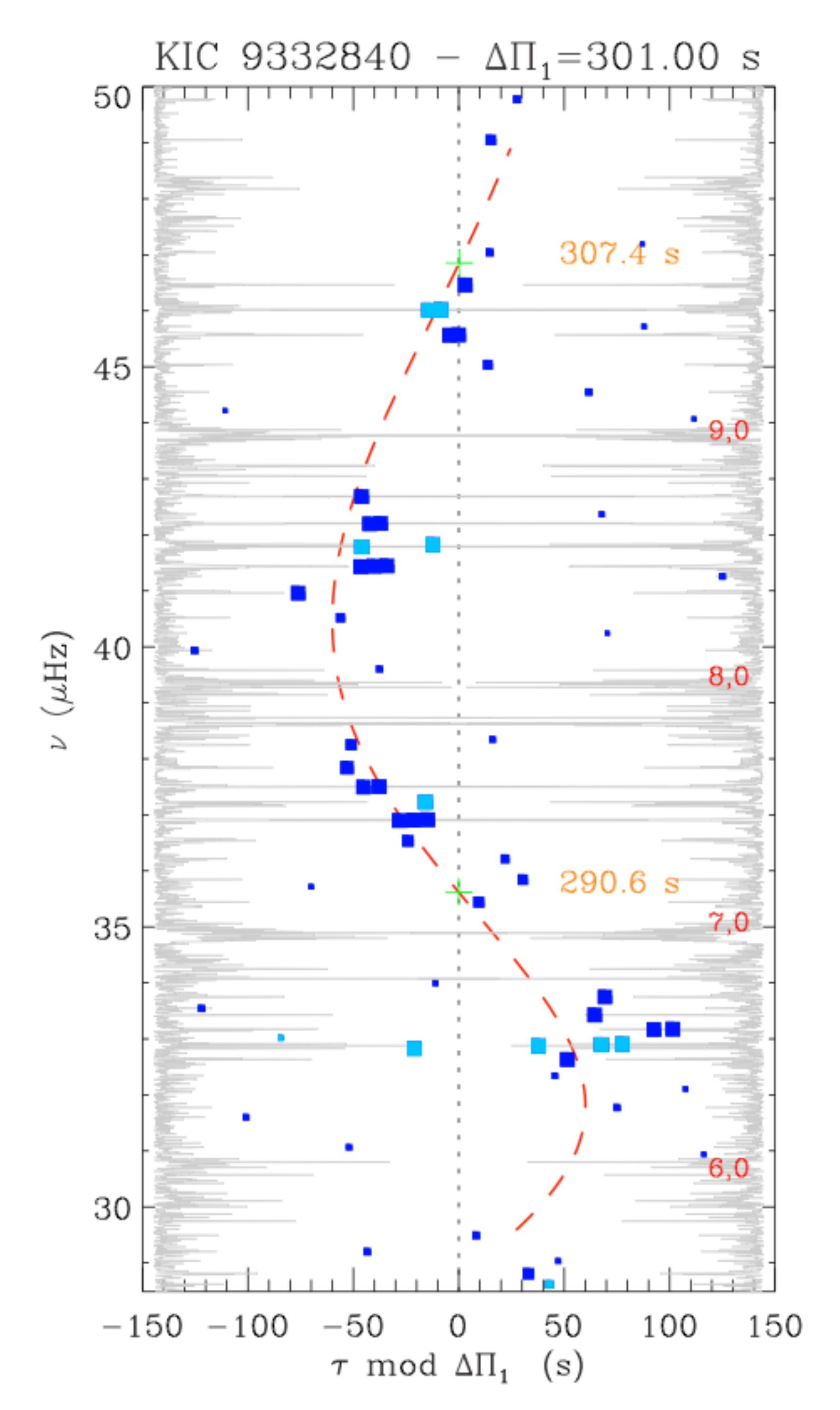}
\caption{Stretched period \'echelle diagram (see explanations in \citealt{mosser15}) of a \kepler\ CHeB giant. The blue squares indicate high peaks in the oscillation spectrum of the star. In the absence of a buoyancy glitch, they are expected to line up on a straight ridge. Here, a modulation is observed (orange dashed line), which is compatible with a buoyancy glitch. Figure from \cite{mosser15}.
\label{fig_mosser15}}
\end{figure}

Further constraints could also be obtained in the near future by detecting the signature of buoyancy glitches in the period spacings of g modes in CHeB giants. The mixing schemes presented in Sect. \ref{sect_core_clump} predict very different abundance profiles in the region above the fully mixed core. For instance, models computed with standard overshooting show step-like features in the helium abundance above the core, while models computed with semi-convection have smooth helium profiles. Sharp variations of $\mu$ are expected to be felt as buoyancy glitches by g modes, which should produce an oscillatory component in the period spacing, as was described in Sect. \ref{sect_gmodes}. The occurrence of buoyancy glitches in the cores of red giants and their seismic signature in the period spacing of g modes has been extensively addressed by \cite{cunha15} using stellar models. Detecting these modulations in $\Delta\Pi_1$ is more complicated for CHeB giants than for main sequence g-mode pulsators because of the mixed character of the modes. Nonetheless, the method of \cite{mosser15} can be used to recover the period spacings of pure gravity modes and thus reveal potential periodic modulations produced by glitches (see Fig. \ref{fig_mosser15}). Glitches produced by sharp $\mu$-gradients above the mixed core are located deep within the g-mode cavity and are thus expected to produce long-period modulations. A systematic search for such features in the oscillation spectra of CHeB giants observed with \kepler\ should bring strong constraints on the way chemical elements are mixed above the convective core.

\section{Conclusion}

The advent of space asteroseismology has yielded numerous novel constraints on the properties of convective cores for stars with various masses and evolutionary stages. We started this review by mentioning that three physical quantities needed to be known to progress in our modeling of the boundary of convective cores. We conclude by summarizing the recent findings of asteroseismology for each of them:
\begin{enumerate}
\item \textit{Distance over which mixed cores are extended:} We here presented only a small selection of all the seismic studies that provided constraints on the extent of the mixed core. The great majority of them concluded that an extension of the mixed core beyond the Schwarzschild limit needed to be invoked. These studies also showed that large star-to-star variations exist for the distance of the extra-mixing. Nevertheless, tendencies can be found in the available data. Main sequence intermediate-mass stars seem to require extensions of the order of 0.2-$0.3\,H_P$. For lower-mass stars ($1.1\lesssim M/M_\odot\lesssim1.5$), lower extensions are needed (from 0.05 to $0.2\,\min(H_P,r_{\rm s}$), where $r_{\rm s}$ is the formal boundary of the convective core). In this mass range, a potential increase of the distance of extra-mixing with stellar mass has been reported but needs to be confirmed. In this review, we have focused on low- and intermediate-mass stars, which have so far benefitted more from space-based asteroseismology, but seismic constraints have also been obtained on the core properties of massive stars. The seismic analyses of $\beta$ Cephei pulsators (8 to 20 $M_\odot$), essentially with ground-based observations, have shown quite large variations in the extent of the extra-mixed region from one star to another, typically ranging from 0 to 0.3 $H_P$ (e.g., \citealt{dupret04}, \citealt{ausseloos04}, \citealt{aerts11}, \citealt{briquet12}). Finally, it has been found that the convective core of core-helium-burning stars needs to be extended over even larger distances, likely in the range of 0.5-$1\,H_P$.
\item \textit{Nature of the mixing in the core extension:} Seismology is currently the only tool to test how efficient the mixing of chemicals is beyond the edge of the convective core. Gravity modes, through their sensitivity to the gradient of $\mu$, are particularly well suited for this purpose. The seismic study of three SPB stars has consistently shown that a diffusive overshooting modeled with an exponentially decaying diffusion coefficient yields better agreement with seismic observations than an instantaneous mixing in the overshoot region. Other constraints on the nature of the mixing could be brought in the near future by using mixed modes in subgiants.
\item \textit{Temperature stratification in the region of extra-mixing:} Measuring this quantity is particularly difficult. However, having penetrative ($\nabla=\nabla_{\rm ad}$) or non-penetrative ($\nabla=\nabla_{\rm rad}$) convection changes the propagation of gravity modes in the overshoot region. This modifies the period spacing of g modes. Hints in favor of non-penetrative convection were obtained from the core-helium burning giants of an old open cluster. Further constraints could be obtained from SPB and $\gamma$ Doradus stars. We here note that constraints have been obtained on the temperature stratification at the bottom of the envelope convection of the Sun. \cite{christensen11} found evidence for a smooth transition from $\nabla=\nabla_{\rm ad}$ to $\nabla=\nabla_{\rm rad}$ in the overshoot region. 
\end{enumerate}

We note that in this review, we have focused exclusively on results obtained with the forward modeling approach. Seismic inversions also have a large potential to bring information on the properties of convective cores. Recent studies have shown promising results for solar-like pulsators (\citealt{bellinger17}, \citealt{buldgen18}) and new, model-independent constraints could come from such analyses in the near future.

The number of targets for which the edge of the mixed core could be seismically probed is increasing rapidly. We are starting to build large enough samples so that trends can be searched in the properties of the extra-mixed region as a function of global stellar parameters. On the short term, this can help us calibrate more refined models of convective core extensions in evolutionary codes. This could provide us with more reliable stellar ages, which is crucial for disciplines that require high-precision stellar modeling, such as the characterization of exoplanets, with the upcoming \plato\ mission, or galactic archaeology. Even more challenging will be the task of disentangling the contributions from the different physical processes to the extensions of convective cores. So far a pragmatic approach has generally been adopted, whereby the effects of all these processes are modeled together in a parametric way. To establish the contribution of rotational mixing, it would be very interesting to search for correlations between the amount of mixing beyond convective cores and the rotational properties of stars.  
%This could provide information about the contribution of rotational mixing to the extension of convective cores. For instance, \cite{moravveji15} and \cite{moravveji16} found the need to include a mild turbulent mixing in the radiative interior, which could be indicative of near solid-body rotation in this region of the star. 
Stars for which seismology can provide measurements of the size of the mixed core and the internal rotation profile would be particularly useful. Magnetic fields are also expected to play a role by inhibiting rotational mixing through the damping of differential rotation in radiative interiors. For instance, this might be happening in the $\beta$ Cephei pulsator V2052 Ophiuci, which hosts a fossil magnetic field with $B_{\rm pol} \sim 400$ G. Through a seismic modeling of the star, \cite{briquet12} found that it indeed has an unexpectedly low amount of extra-mixing beyond the convective core. More studies of this type are needed to progress in our understanding of the processes that can extend the size of mixed cores. In this context, the \textsc{TESS} and \plato\ missions are particularly welcome. They will provide us with seismic data with a nearly all-sky coverage, which will greatly increase the number of targets for which seismic constraints on the core properties can be derived. In particular, with \plato\ data, we will be able to perform much more meaningful statistical studies of the extent of the mixed core in solar-like pulsators.

%\begin{table}
%\caption{This is an example of table.
%\label{table1}} 
%\small
%\begin{center} 
%\begin{tabular}{| l | r r r |}
%\hline 
% & Col.1 & Col.2 & Col.3 \\
%\hline
%Line 1 & ... & ... & ... \\
%Line 2 & ... & ... & ... \\
%\hline 
%\end{tabular} 
%\end{center} 
%\end{table} 

%\begin{figure}[h]
%\centering
%\includegraphics[width=8cm]{fig1.pdf}
%\caption{This is an example of a single figure. \label{fig_1}}
%\end{figure}

%\begin{figure}[h]
%\begin{minipage}{8cm}
%\centering
%\includegraphics[width=8cm]{fig2.pdf}
%\caption{This is the caption of the left-hand side figure. \label{fig_2}}
%\end{minipage}
%\hfill
%\begin{minipage}{8cm}
%\centering
%\includegraphics[width=8cm]{fig3.pdf}
%\caption{This is the caption of the right-hand side figure. \label{fig_3}}
%\end{minipage}
%\end{figure}

%
% USE A SECTION WITHOUT NUMBER FOR THE ACKNOWLEDGEMENTS
%
\section*{Acknowledgements}
I am thankful to Marc-Antoine Dupret for inviting me to the Li\`ege Workshop organized in honor of Arlette Noels. I also take the opportunity to express my deep gratitude to Arlette, for all the very enlightening discussions that I have had with her. She is a great source of inspiration for me. I also acknowledge support from the project BEAMING ANR-18-CE31-0001 of the French National Research Agency (ANR) and from the Centre National d'Etudes Spatiales (CNES).
%
% BEGIN THE REFERENCE LIST WITH \beginrefer
% USE \refer BEFORE THE REFERENCES AND BEGIN A NEW PARAGRAPH AFTER THE 
% REFERENCE !
% DO NOT FORGET TO END THE LIST WITH \endrefer
% 
%
% INSTRUCTIONS FOR BIBLIOGRAPHY:
% ==============================
% - DON'T USE THE & SYMBOL
% - USE INITIALS FOR FIRST AND MIDDLE NAMES, AND SPECIFY FULL FAMILY NAME (see examples below)
% - NO COMMA BETWEEN NAME AND INITIALS
% - USE COMMA BETWEEN DIFFERENT AUTHORS NAMES
% - NO COMMA AFTER THE LAST AUTHOR NAME
% - FOR LONG AUTHOR LISTS, SPECIFY THE FIRST 3 AUTHORS FOLLOWED BY 'et al.', WITH NO COMMA BEFORE AND AFTER 'et al.'
% - INSERT A BLANK SPACE BETWEEN MULTIPLE INITIALS
% - USE STANDARD JOURNAL ACRONYMS FREQUENTLY USED IN MAIN ASTROPHYSICS JOURNAL
% - SORT REFERENCES BY ALPHABETICAL ORDER OF FIRST AUTHOR NAMES
% - MULTIPLE REFERENCES WITH THE SAME FIRST AUTHOR SHOULD BE SORTED BY CHRONOLOGICAL ORDER

\footnotesize
\bibliographystyle{aa.bst} % style aa.bst
\bibliography{biblio} % your references Yourfile.bib

\end{document}